\documentclass[preprint,12pt,nopreprintline]{elsarticle}

\usepackage{tabularx}

\usepackage{amsmath,amsfonts}
\usepackage{algorithmic}
\usepackage{algorithm}
\usepackage{array}
\usepackage{float}
\usepackage{textcomp}
\usepackage{stfloats}
\usepackage{url}
\usepackage{verbatim}
\usepackage{graphicx}
\usepackage{amsmath, xparse}
\usepackage[utf8]{inputenc}
\usepackage{xcolor}
\usepackage{amsmath,amssymb,amsfonts}
\usepackage{algorithmic}
\usepackage{graphicx}
\usepackage{textcomp}
\usepackage{bm}
\usepackage{float}
\usepackage{xcolor}
\usepackage{soul}
\usepackage{xfrac}
\usepackage{subfigure}

\definecolor{LightGray}{gray}{0.8}
\definecolor{Orange}{rgb}{1.0, 0.31, 0.0}
\definecolor{Green}{rgb}{0.3, 1.0, 0.3}
\definecolor{Blue}{rgb}{0.75,0.75,1}

\usepackage{amssymb}

\journal{}
\biboptions{sort&compress}
\begin{document}

\begin{frontmatter}



\title{A Time-Domain Method of Auxiliary Sources for Efficient Analysis of Transient Electromagnetic Scattering by Moderately Conductive Cylinders}


\author[label1]{Minas Kouroublakis}

\author[label1]{Nikolaos~L.~Tsitsas }

\author[label2]{Yehuda Leviatan}

\affiliation[label1]{organization={School of Informatics, Aristotle University of Thessaloniki},
            city={Thessaloniki},
            postcode={54124}, 
            country={Greece}}

\affiliation[label2]{organization={Department of Electrical Engineering, Technion-Israel Institute of Technology},
            city={Haifa},
            postcode={32000}, 
            country={Israel}}

\begin{abstract}
This paper presents a time-domain implementation of the Method of Auxiliary Sources (MAS) combined with the Standard Impedance Boundary Condition (SIBC) for electromagnetic scattering problems involving cylindrical scatterers with finite but moderate conductivity. The proposed approach focuses on solving the two-dimensional problem using a first-order SIBC, which is valid when the conductivity is sufficiently higher than the maximum spectral frequency times the dielectric permittivity of the scatterer. This regime includes moderately conductive materials--such as carbon-based composites, conductive polymers, and doped dielectrics--that are increasingly used in real-world radio-frequency applications, including wearable electronics, electromagnetic interference shielding, and biomedical sensors. Under the above validity conditions, the interaction between the incident wave and the scatterer is dominated by surface effects, allowing for an efficient and accurate modeling strategy without the need to compute internal fields. The theoretical formulation of the time-domain MAS-SIBC method is developed, followed by extensive numerical testing on various geometries whose cross section is a closed curve. Such geometries include circular, elliptical, super-circular, rounded-triangular, and inverted-elliptical scatterers. A planar geometry is also tested. All results are validated against analytical solutions and commercial frequency-domain solvers, demonstrating the accuracy and practical potential of the proposed method. The findings suggest that time-domain MAS-SIBC offers a promising and computationally efficient approach for modeling scattering from materials even with moderate conductivity.
\end{abstract}




\end{frontmatter}


\section{Introduction}
\label{sec:Intro}
Time-domain electromagnetic scattering has become increasingly important thanks to its wide range of applications, spanning from 
radars \cite{bennett1978timedomain, luo2021timedomain,muppala2024fast,mu2021joint}, geophysical exploration \cite{carrasco2022time, yang2012three, parshin2021lightweight}, and non-destructive testing \cite{jawad2021ifft, guo2022review} to electromagnetic shielding studies \cite{celozzi2022electromagnetic, lovat2023transient,yan2021shielding} and biomedical imaging \cite{wang2021overview, saraskanroud2021comparison}. Unlike frequency-domain approaches, which focus on how a system responds at specific frequencies, time-domain methods use short electromagnetic pulses to probe a much broader range of frequencies at once. The width of the temporal pulse determines how wide this frequency range is--the shorter the pulse, the broader the spectrum it covers. This ability to capture rich, broadband information makes time-domain techniques a powerful tool for studying complex materials and structures, detecting hidden objects, exploring the subsurface, and ensuring electronic systems perform reliably in fast-changing environments.

Several numerical methods have been developed over the years to tackle time-domain electromagnetic scattering problems \cite{ren2022advances}. Among them, the Finite-Difference Time-Domain (FDTD) method is by far the most widely used, thanks to its flexibility and straightforward implementation \cite{gedney2022introduction,teixeira2023finite, mccoy2021finite}. A Time Domain-Finite Element Method (TD-FEM) has also been implemented \cite{yan2015theoretical} and used in commercial software such as COMSOL Multiphysics \cite{Comsol}. 
Galerkin time domain methods have been developed through the years \cite{alvarez2012spurious,angulo2015discontinuous}. 
Surface- and volume-integral-equation based methods are other popular choices  \cite{ takahashi2023fast, chen2020explicit, sayed2015stable}. All these techniques rely on meshing either the volume or the surface of the scatterer. On the other hand, the Method of Auxiliary Sources (MAS) \cite{papakanellos2024method}, also known as the Method of Fundamental Solutions (MFS) \cite{Karageorghis2025, Cheng2020, Barnett2008}, and the Source Model Technique (SMT) \cite{leviatan1988generalized, tsitsas2018}, offers a mesh-less alternative together with other methods \cite{hochman2013use} related to the Generalized Multipole Technique \cite{6095683}. MAS has already shown solid results in certain time domain two-dimensional scattering problems \cite{ludwig2006towards, ludwig2011source, ludwig2007source, ludwig2008time} and is considered a promising approach for more complex scenarios, due to its efficiency and potential to handle intricate geometries without the need for traditional meshing.

When dealing with scatterers characterized by moderate to high conductivity \(\sigma\), the SIBC \cite{senior1960impedance} proves to be a particularly effective tool. Rather than solving for the electromagnetic fields both inside and outside the scatterer, the SIBC allows for the direct modeling of the scattered field, significantly reducing computational costs. 
This makes it particularly useful in cases where the internal fields are not of primary interest. 
In the frequency domain, the SIBC has been successfully integrated into a variety of numerical methods \cite{senior1995approximate,  anastassiuSIBC2002, anastassiuSIBC2005, hoppe2018impedance}. In the time domain, numerous works have explored its application within FDTD \cite{maloney1992use, oh1995efficient, kobidze2010implementation, mak2013fdtd, karkkainen2003fdtd}, among others. Additionally, several studies have investigated the TD-FEM method \cite{celozzi1993time, sabariego2012time, tsakanian2015broadband}. A time-domain SIBC has also been implemented within the BEM framework \cite{yuferev1998efficient, yuferev2000invariant, yuferev1999unified, di2008computation}. Both first-order and higher-order versions of the SIBC have been developed in these methods. However, a closer examination of the literature reveals that most studies focus on relatively simple scatterer geometries, such as planar surfaces or fundamental closed curves like rectangles, circles, and ellipses. 

In this paper, we implement the time-domain SIBC within the framework of MAS, thereby paving the way for more efficient and accurate modeling of complex scattering problems. The combined time-domain SIBC-MAS approach is applied for cylindrical scatterers, the cross section of which can be in general described by an arbitrary smooth curve. We focus on using a first-order SIBC, which is known to be valid for scatterers with sufficiently high conductivity. More precisely, the conductivity $\sigma$ must satisfy the condition $\sigma \gg \omega_{\text{max}} \varepsilon$, where $\omega_{\text{max}}$ is the highest frequency present in the spectrum of the incident pulse and $\varepsilon$ is the dielectric permittivity of the scatterer. Materials that meet this condition include not only traditional metals but also a wide range of moderately conductive substances with conductivities in the range of 5--15~S/m at radio frequencies. Examples include carbon-based composites \cite{feng2024conductivity,zhao2019review}, conductive polymers \cite{takeda2011modeling, kunz2019direction}, and doped dielectrics \cite{mamin2001conductivity}, all of which are used in real-world radio-frequency applications such as flexible electronics, electromagnetic interference shielding, and biomedical sensing. Under the above condition, and in accordance with Snell's law, the transmitted wave inside the scatterer propagates almost normally to the local surface and attenuates rapidly, with minimal lateral diffusion. Consequently, the interaction between the incident wave and the scatterer is dominated by surface effects, allowing accurate modeling without resolving the internal field distribution. This formulation enables the SIBC-MAS framework to be extended beyond simple planar surfaces and applied to cylinders whose cross sections boundaries are closed curves with both concave and convex geometrical features.

The paper is organized as follows. In Section II, we present the theoretical framework for implementing MAS in combination with SIBC in the TD. Section III is dedicated to numerical results, where we apply the developed method to a variety of scatterers, including circular, elliptical, super-circular, rounded-triangular, and inverted-elliptical geometries. For completeness, we also consider a planar geometry to further test the capabilities of the developed approach. All numerical results are validated through comparison with available analytical solutions and commercial frequency-domain solvers. Finally, Section IV summarizes the main findings of the work and discusses potential directions for future research.

\section{Scattering by a Conductive Cylinder: Time-Domain Analysis and Implementation of the SIBC}
\label{sec:theory}
\subsection{Frequency-Domain Analysis of the SIBC}
\begin{figure} [htb!]
\centering
\includegraphics[scale=0.31]{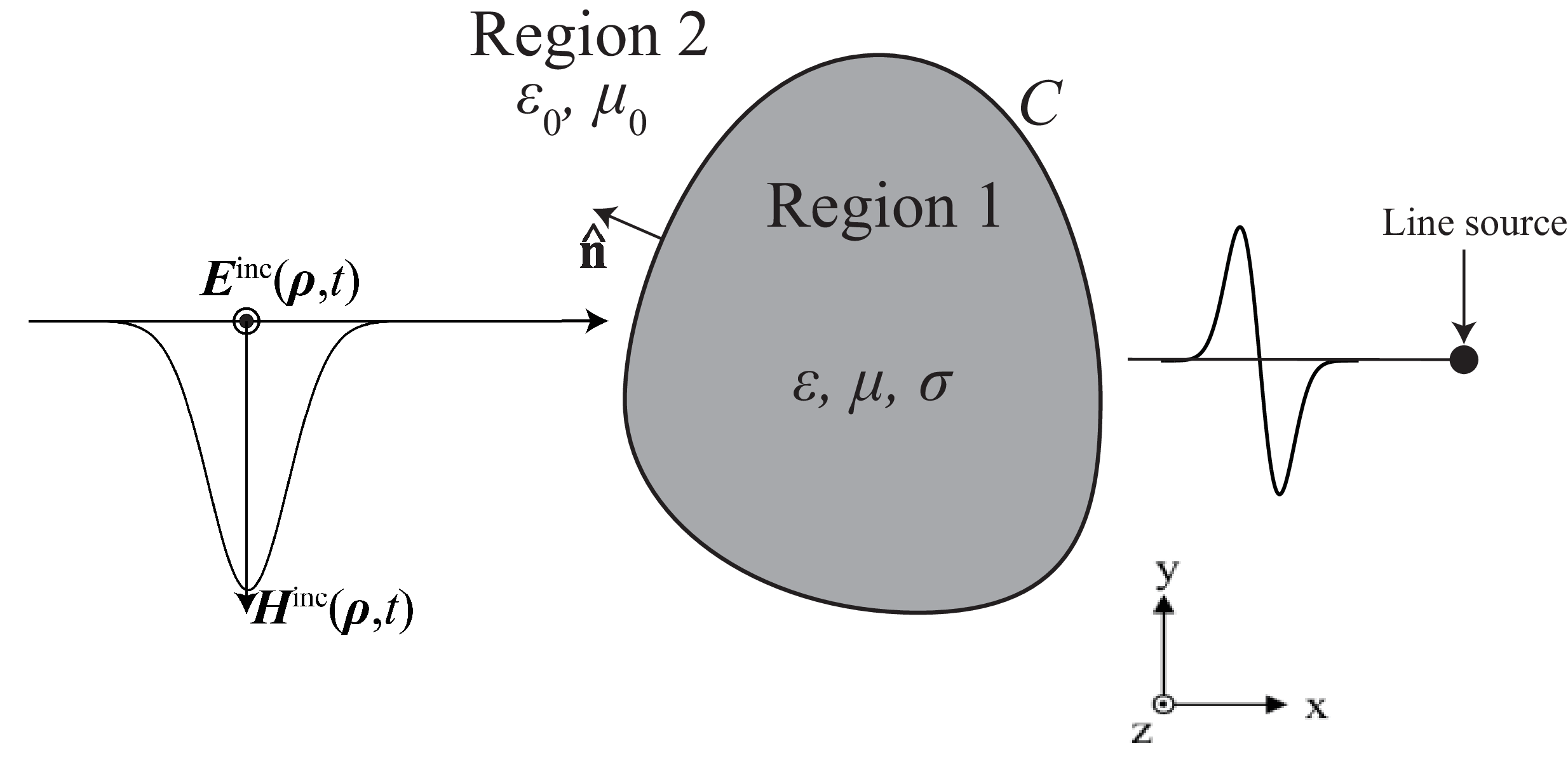}
\caption{Schematic of the considered two-dimensional conductor illuminated by a transient plane wave or by a transient cylindrical wave originating from a line source.}
\label{fig:original}
\end{figure}

A cylinder of infinite length along the $z$-axis, with a smooth arbitrary cross-sectional shape, is shown in Fig.~\ref{fig:original}. The cylinder is assumed to be non-magnetic, with finite conductivity $\sigma$ and dielectric permittivity $\varepsilon$, and is surrounded by vacuum with dielectric permittivity $\varepsilon_0$ and magnetic permeability $\mu_0$. The region outside (inside) the scatterer is denoted by $R_2$ ($R_1$). The cylinder is illuminated by a TM\(_z\) or TE\(_z\) wave, which may be either a plane wave, or a cylindrical wave originating from an infinitely long filament of electric ($\rm TM_z$) or magnetic ($\rm TE_z$) current located near the scatterer.

One method for solving the formulated scattering problem is the application of the surface equivalence principle (SEP), which states that the scattered and internal fields can be represented in terms of surface electric and magnetic current densities on the boundary $C$ of the cross section of the cylindrical scatterer. By enforcing the boundary conditions that ensure the continuity of the tangential components of the electric and magnetic fields on $C$, a corresponding system of integral equations is derived. Solving this system yields the unknown current densities and ultimately leads to the unique determination of the fields in regions $R_1$ and $R_2$.

In problems where the scatterer is homogeneous, such as the one of Fig.~\ref{fig:original}, usually the primary focus is on the computation of the scattered field in $R_2$. In such cases, rather than enforcing both boundary conditions for the continuity of the tangential electric and magnetic fields on $C$, the Standard Impedance Boundary Condition (SIBC), also known as the Leontovich boundary condition, is applied. This condition is still utilized within the framework of SEP, replacing the exact boundary conditions with an approximate relation between the equivalent surface currents and the tangential fields on the scatterer's surface. A key advantage of this approach is that it significantly reduces computational complexity compared to solving the full system of integral equations, required in the general case. 
Thus, it is particularly useful for large or moderate to high-frequency scattering problems. In the frequency domain (FD), assuming a time dependence of the form $\exp(\mathrm i\omega t)$, the SIBC is expressed by
\begin{equation}
\mathbf E_{\rm tan}(\boldsymbol \rho,\omega)=Z_s(\omega)\,\mathbf {\hat n}\times \mathbf H_{\rm tan}(\boldsymbol \rho,\omega)
\label{eq:leontovich_fd}
\end{equation}
where $\mathbf E_{\rm tan}$ ($\mathbf {H_{\rm tan }}$) is the total tangential electric (magnetic) field (incident and scattered) on the boundary $C$, $\boldsymbol \rho$ is the position vector of a boundary point, $\omega$ is the angular frequency, $\mathbf {\hat n}$ is the outward normal unit vector on $C$, and $Z_s$ is the frequency dependent surface impedance of the scatterer.



In case where $\sigma \gg \omega \varepsilon$, the surface impedance is given by \cite{balanis2012}
\begin{equation}
Z_s (\omega)= \sqrt{\frac{\mathrm{i} \omega \mu_0}{\sigma}},
\end{equation}
%
%
%
%
and, from (\ref{eq:leontovich_fd}), we get
\begin{equation}
\mathbf E_{\rm tan}(\boldsymbol\rho,\omega)
=\zeta_s(\omega)\,\mathbf {\hat n}  \times(\mathrm i \omega \mathbf H_{\rm tan}(\boldsymbol \rho,\omega)),
\label{eq:SIBC_high}
\end{equation}
where
\begin{equation}
\zeta_s(\omega)=\sqrt{\frac{\mu_0}{\mathrm i \omega \sigma}}.
\label{eq:zeta_w}
\end{equation}

This formulation will facilitate the derivation of the TD SIBC in the subsequent analysis.

\subsection{Time-Domain Transformation of the SIBC}
The transition of the SIBC from FD to TD requires the application of the Inverse Fourier Transform (IFT) to (\ref{eq:SIBC_high}). At this point, we keep in mind the well-known property that the product of two functions in FD transforms into the convolution of the IFTs of these functions in TD.
Applying IFT on (\ref{eq:SIBC_high})  and taking into account that $\mathrm i \omega\, \mathbf {\hat n} \times \mathbf H_{\rm tan}(\boldsymbol \rho, \omega) $ transforms to $\mathbf {\hat n}\times (\partial \mathbf H_{\rm tan}(\boldsymbol \rho,t) /\partial t)$, we obtain
\begin{equation}
\boldsymbol  E_{\rm tan}(\boldsymbol \rho,t)=\int_{-\infty}^{+\infty}\zeta_s(\tau)\,\mathbf {\hat n}\times \frac{\partial \boldsymbol H_{\rm  tan}(\boldsymbol \rho,t-\tau)}{\partial t}\mathrm{d}\tau,
\label{eq:Tesche}
\end{equation}
where \cite{tesche1990inclusion}
\begin{equation}
\zeta_s(t)=\sqrt{\frac{\mu_0}{\pi\sigma t}}
\end{equation}

By assuming causality ($t>0$) and using commutativity, (\ref{eq:Tesche}) is written as

\begin{equation}
\boldsymbol  E_{\rm tan}(\boldsymbol \rho,t)=\int_{0}^{t}\zeta_s(t-\tau)\,\mathbf {\hat n}\times \frac{\partial \boldsymbol H_{\rm  tan}(\boldsymbol \rho,\tau)}{\partial t}\mathrm{d}\tau
\label{eq:Tesche_2}
\end{equation}









\subsection{Time-Domain MAS and Implementation of the SIBC}
\begin{figure} [htb!]
\centering
\includegraphics[scale=0.31]{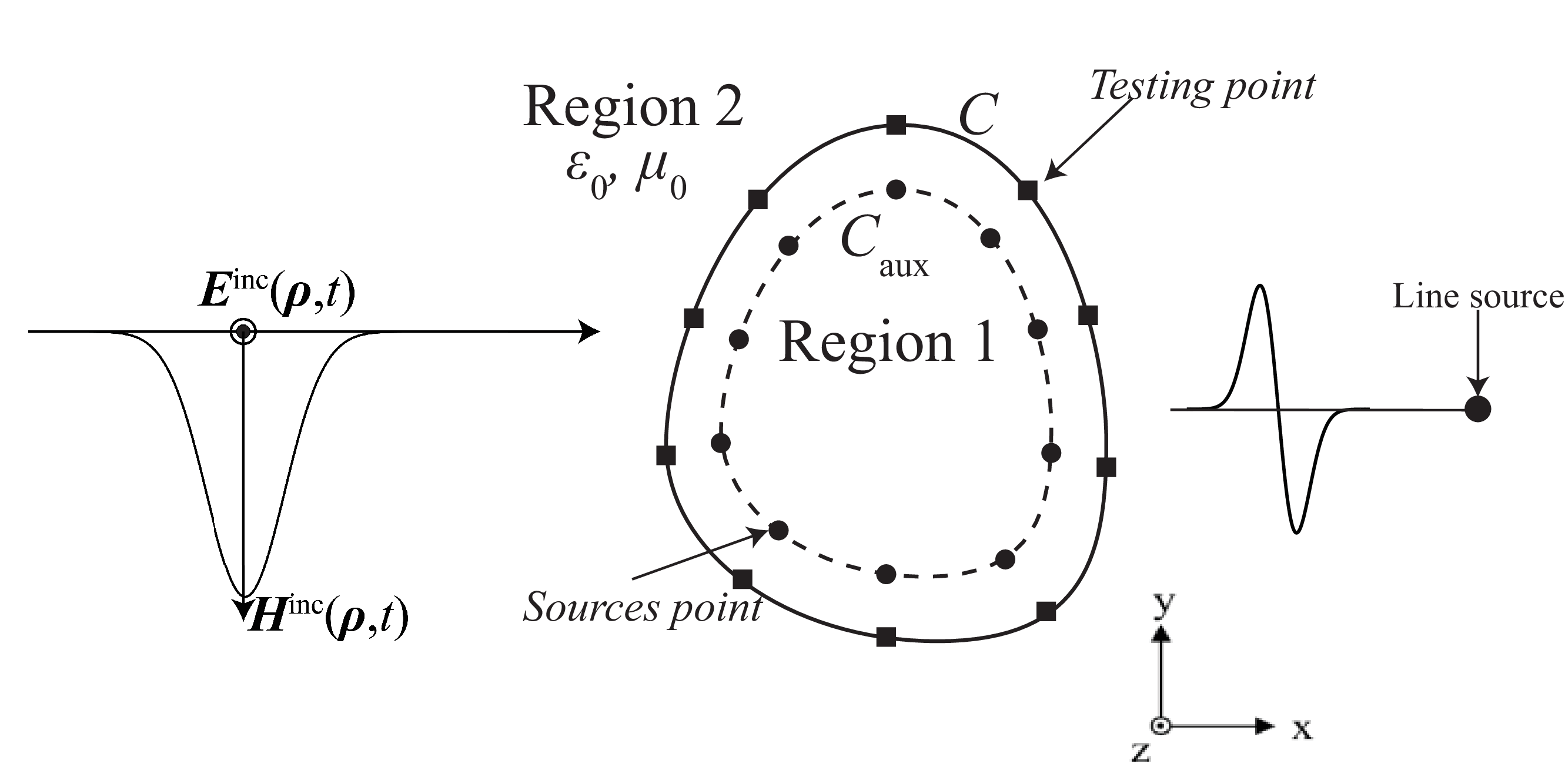}
\caption{Simulated equivalence in the external region $R_2$, established by the fictitious combined sources and evaluated at the testing points on $C$.}
\label{fig:MAS_appl}
\end{figure}

To solve the time-dependent integral equation  (\ref{eq:Tesche_2}), we will use the SEP in the framework of MAS as shown in Fig. \ref{fig:MAS_appl} and applied in \cite{ludwig2006towards}. 
From this point onward, the analysis will focus on the \( \rm TM_z \) polarization. A corresponding solution for the case of $\rm TE_z$ incidence can be easily obtained from this formulation by swapping the electric and magnetic sources.

Unlike the conventional approach where the equivalent current densities are defined on $C$, in MAS, the equivalent currents are placed on an auxiliary surface $C_{\rm aux}$ 
located inside region $R_1$, and, thus, displaced from $C$. 
The current densities may either be electric $\mathbf J_{s_i}$ or magnetic $\mathbf M_{s_i}$ ones. 
Within the combined sources (CS) formulation \cite{ludwig2006towards}, a combination of both electric and magnetic currents is employed, both stemming from a common parent electric current distribution $\mathbf {\tilde J}_{s_i}$. For $\rm TM_z$ polarization, we have  
\begin{equation}
\left( \mathbf J_{s_i}, \mathbf M_{s_i} \right)=\left( (1-\alpha)\mathbf {\tilde{J}}_{s_i}, \alpha \eta\,\mathbf {\hat n_{\rm aux}}\times \mathbf{\tilde{J}}_{s_i} \right),~~0\le\alpha\le1,
\label{eq:densities}
\end{equation}
where $\alpha$ is a constant that governs the relative contributions of the electric and magnetic components in the total current distribution, \( \eta = \sqrt{\mu_0 / \varepsilon_0} \) is the impedance of $R_2$, and \( \mathbf{\hat{n}_{\rm aux}} \) is the outward normal unit vector on \( C_{\rm aux} \). From (\ref{eq:densities}), it is inferred that when $\alpha = 0$ \, ($\alpha = 1$)
only the electric (the magnetic) fictitious currents are present.

The scattered electric and magnetic fields radiated by the current densities are given by the following integrals \cite{ludwig2006towards}
\begin{multline}
\boldsymbol E^{\rm sc}_z(\boldsymbol \rho, t)=-\frac{\eta}{2\pi}\int_{C_{\rm aux}}^{}\int_{-\infty}^{t-P/c}\frac{1}{\sqrt{c^2(t-t')^2-P^2}} \\ \left\{  \frac{\partial J_{s_{i},z}(\boldsymbol \rho', t')}{\partial t'}+\frac{\mathbf {\hat z}}{\eta}\cdot\left[ \left( \frac{\mathbf M_{s_i}(\boldsymbol\rho', t')}{(t-t')}+\frac{\partial \mathbf M_{s_i}(\boldsymbol\rho', t') }{\partial t'} \right)\times\frac{\mathbf P}{c(t-t')} \right]\right\}\mathrm{d}t' \mathrm{d}l'
\label{eq:cont_electric}
\end{multline}
\begin{multline}
\boldsymbol H^{\rm sc}(\boldsymbol \rho, t) = \frac{c}{2\pi} \int_{C_{\rm aux}} \int_{-\infty}^{t-P/c} 
\frac{1}{\sqrt{c^2(t - t')^2 - P^2}} \\
\left\{ \mathbf{\hat z} \times \frac{\mathbf{\hat P}}{P} (t - t') 
\frac{\partial J_{s_i,z}(\boldsymbol \rho', t')}{\partial t'} + 
\frac{2c^2(t - t')^2 - P^2}{2\eta c P^2} \right. \\
\left. \left[ \left( \frac{\partial \mathbf M_{s_i}(\boldsymbol \rho', t')}{\partial t'} \cdot \mathbf{\hat P} \right)\mathbf{\hat P} +
\left( \frac{\partial \mathbf M_{s_i}(\boldsymbol \rho', t')}{\partial t'} \times \mathbf{\hat P} \right) \times \mathbf{\hat P} \right] \right. \\ \left. 
- \frac{1}{2\eta c} \frac{\partial \mathbf M_{s_i}(\boldsymbol \rho', t')}{\partial t'} \right\} \mathrm{d}t' \mathrm{d}l'
\label{eq:cont_magne}
\end{multline}
where $c$ is the light speed in vacuum, $\boldsymbol \rho$ is the observation position vector in $R_2$,  $\boldsymbol \rho'$ is the position vector on $C_{\rm aux}$, while $\mathbf P=\boldsymbol \rho-\boldsymbol \rho'$, and $P=||\mathbf P||$.


Next, the fictitious current densities \( \mathbf J_{s_i} \) and \( \mathbf M_{s_i} \) are  represented as superpositions of \( N \) pairs of electric and magnetic filamentary current sources, each positioned at specific locations \( \boldsymbol {\rho'_n} \) on \( C_{\rm aux} \). 
For \( \rm{TM}_z \) (\( \rm{TE}_z \)) polarization, the electric (magnetic) current source corresponds to a \( z \)-directed electric (magnetic) filament, whereas the magnetic (electric) current source is modeled as a \( z \)-directed filament comprising an array of infinitesimal magnetic dipoles, aligned tangentially to \( C_{\rm aux} \).

Applying a straightforward temporal discretization, we obtain that the time derivative of the parent current surface density $\mathbf{\tilde{J}}_{s_i}$ is expressed as
\begin{equation}
\frac{\partial {\tilde J}_s(\boldsymbol \rho', t')}{\partial t'}=\sum_{n}^{}\sum_{k'}^{}\tilde I_n^{(k')}\delta _s(\boldsymbol \rho'-\boldsymbol \rho_n')T_n^{(k')}(t')
\end{equation}
where \( \tilde{I}_n^{(k')} \) represents an unknown amplitude associated with the \( n \)-th combined source at the \( k' \)-th time step, while \( \delta_s \) denotes the surface delta function, and \( T_n^{(k')}(t') \) is the simplest temporal basis function, defined as
\begin{equation}
T_n^{(k')}(t')=\begin{cases}
1,~k'\Delta t<t'+\frac{ P_{\tilde m_n,n}}{c}<(k'+1)\Delta t\\
0,~~~\text {otherwise}
\end{cases}.
\end{equation}
In this context, \( P_{m,n} = |\boldsymbol \rho_m - \boldsymbol \rho'_n| \) represents the distance between the \( n \)-th source point \( \boldsymbol \rho'_n \) and the \( m \)-th testing point \(\boldsymbol \rho_m \) among the \( M \) testing points on $C$ where the SIBC will be applied. The index \( \tilde{m}_n \) is defined as the testing point closest to the \( n \)-th source, ensuring that \( P_{\tilde{m}_n, n} = |\boldsymbol \rho_{\tilde{m}_n} - \boldsymbol \rho'_n| \) corresponds to the shortest distance between the \( n \)-th source and any of the testing points on the cylinder's boundary. Since the fictitious filamentary sources are positioned on a surface retracted by a certain distance from the cylinder’s boundary, the retarded temporal discretization starts at \( t' = -P_{\tilde{m}_n,n}/c \). This retarded time of the sources is permitted to take negative values, ensuring that the fields radiated by the filamentary sources reach the surface precisely when the incident wave arrives, i.e., at \( t = 0 \).
Using the relationship between electric and magnetic currents given in (\ref{eq:densities}), expressions for the discretized time derivatives of the fictitious electric and magnetic surface current densities are derived as follows:
\begin{equation}
\begin{cases}
\frac{\partial J_{s,z}(\boldsymbol{\rho}', t') }{\partial t'}=\sum_{n}^{}\sum_{k'}^{}(1-\alpha)\tilde I_n^{(k')}\delta_s(\boldsymbol{\rho}'- \boldsymbol{\rho}'_n)T_n^{(k')}(t')\\
\frac{\partial M_{s,z}(\boldsymbol{\rho}', t') }{\partial t'}=\sum_{n}^{}\sum_{k'}^{}\alpha \eta\tilde I_n^{(k')}\mathbf{\hat l}_n\delta_s(\boldsymbol{\rho}'- \boldsymbol{\rho}'_n)T_n^{(k')}(t')
\end{cases}
\label{eq:derivatives}
\end{equation}
where $\mathbf{\hat{l}}_n = \mathbf{\hat{n}}_n \times \mathbf{\hat{z}}$ denotes a unit vector tangent to the curve $C_{\text{aux}}$ at the point $\boldsymbol{\rho}'_n$, with $\mathbf{\hat{n}}_n$ being the outward normal unit vector to $C_{\text{aux}}$ at that location.

By inserting (\ref{eq:derivatives}) into (\ref{eq:cont_electric}) and (\ref{eq:cont_magne}) and evaluating at discrete time intervals \( t_i = (i+1)\Delta t \) $(i=0,1,\ldots)$ across the \( M \) testing points \( \boldsymbol{\rho}_m \), we obtain the following \( M \) equations for the electric field at each time step \( i \)
\begin{multline}
\boldsymbol  E_z^{\rm sc}(\boldsymbol \rho_m,(i+1)\Delta t)= -\frac{\eta}{2\pi c}\sum_{n}^{}\Biggl( \sum_{k'=0}^{i-\left\lfloor \kappa_{m,n}  \right\rfloor-1}\tilde I_n^{(k')}\cdot 
[W^{E}_{m,n}(i-k')-W^{E}_{m,n}(i-k'-1)]\\
+\tilde I_n^{(i-\lfloor\kappa_{m,n}\rfloor)}\cdot W^{E}_{m,n}(\lfloor \kappa_{m,n}\rfloor)\Biggr),~m=1,2,\ldots,M 
\label{eq:discrete_electric}
\end{multline}
where $\kappa_{m,n}=(P_{m,n}-P_{\tilde m_n,n})/(c\Delta t)$, and $\lfloor \kappa_{m,n}\rfloor$ denotes the greatest integer less than or equal to $\kappa_{m,n}$. The function $W^{E}_{m,n}$ is expressed as follows (with respect to the dummy variable $x$):
\begin{multline}
W^{E}_{m,n}(x)=(1-\alpha)\,\text{acosh}\left( \frac{c\Delta t}{P_{m,n}}\left[x+\frac{P_{\tilde m_n,n}}{c\Delta t} \right]+1 \right) \\
+\alpha \mathbf {\hat z}\cdot(\mathbf{\hat l_n}\times \mathbf {\hat P}_{m,n})\sqrt{\left( \frac{c\Delta t}{P_{m,n}}\left[x+\frac{P_{\tilde m_n,n}}{c\Delta t} \right]+1 \right)^2-1}
\end{multline}
where $\mathbf {\hat P}_{m,n}=(\boldsymbol \rho_m-\boldsymbol \rho'_n)/P_{m,n}$. The scattered magnetic field is given by
\begin{multline}
\boldsymbol H^{\rm sc}(\boldsymbol \rho_m, (i+1)\Delta t)= \frac{1}{2\pi c}\sum_{n}^{}\bigg( \sum_{k'=0}^{i-\lfloor \kappa_{m,n}\rfloor-1} \tilde I^{(k')}_n \left[\mathbf W^{H}_{m,n}(i-k')-\mathbf W^{H}_{m,n}(i-k'-1)\right] \\
+\tilde I_n^{(i-\lfloor \kappa_{m,n}\rfloor)}\mathbf W_{m,n}^{{H}}(\lfloor\kappa_{m,n}\rfloor)\bigg),~m=1,2,\ldots,M
\label{eq:discrete_magnetic}
\end{multline}
with
\begin{multline}
\mathbf W_{m,n}^{H}(x)=\bigg\{ \left[ (1-\alpha)\mathbf{\hat z}\times \mathbf{\hat P_n}+\frac{\alpha}{2}\mathbf {\hat l'}_n\left( \frac{c\Delta t}{P_n}\left[ x+1+\frac{P_{\tilde m_n,n}}{c\Delta t} \right] \right) \right] \\
\times \left( \frac{c\Delta t}{P_{m,n}}\left[x+\frac{P_{\tilde m_n,n}}{c\Delta t} \right]+1 \right) \bigg\} - \frac{\alpha}{2}\mathbf {\hat l}_n\, \text{acosh}\left( \frac{c\Delta t}{P_{m,n}}\left[x+\frac{P_{\tilde m_n,n}}{c\Delta t} \right]+1 \right)
\end{multline}
We continue the discretization process by applying the SIBC (\ref{eq:Tesche_2}) on the $M$ testing points and assuming the fields are piecewise linear in time, yielding
%
%
%
\begin{multline}
\boldsymbol E_{\rm tan}\Bigl(\boldsymbol \rho_m, (i+1)\Delta t\Bigr)=\sum_{k'=0}^{i}\zeta_s\Bigl((i-k')\Delta t\Bigr)\cdot \\\mathbf {\hat n}_m\times  \Bigl [\boldsymbol H_{\rm tan}\Bigl(\boldsymbol \rho_m,(k'+1)\Delta t\Bigr)-\boldsymbol H_{\rm tan}\Bigl(\boldsymbol \rho_m,k'\Delta t\Bigr)\Bigr],~m=1,2,\ldots,M,
\label{eq:SIBC_high_disc}
\end{multline}
where we used the backward discrete derivative definition. The discretized SIBC  (\ref{eq:SIBC_high_disc}) is expressed in a matrix-type form as follows:
%
%
\begin{multline}
\mathbf V^{(i)}_E+\sum_{k'=0}^{i}[\mathbf Z_E^{(i-k')}]\boldsymbol I^{(k')}=\sum_{k'=0}^{i}\zeta_s\Bigl ((i-k')\Delta t\Bigr)\\\left[ (\mathbf V^{(k')}_H-\mathbf V_H^{(k'-1)})+\sum_{q=0}^{k'}\left( [\mathbf Z_H^{(k'-q)}]\boldsymbol {\tilde I}^{(q)}-[\mathbf Z_H^{(k'-q-1)}]\boldsymbol {\tilde I}^{(q-1)} \right) \right]
\label{eq:generalized_2}
\end{multline}
where 
\begin{equation}
\left[ \mathbf Z_{\scriptsize\left\{ \begin{smallmatrix} E \\ H \end{smallmatrix} \right\}}^{(j)} \right]_{m,n}
= \frac{1}{2\pi c}
\begin{Bmatrix}
-\eta \\
1
\end{Bmatrix}
\begin{cases}
W^{\scriptsize\left\{ \begin{smallmatrix} E \\ H \end{smallmatrix} \right\}}_{m,n}(j) - W^{\scriptsize\left\{ \begin{smallmatrix} E \\ H \end{smallmatrix} \right\}}_{m,n}(j-1), & j > \left\lfloor \kappa_{m,n} \right\rfloor \\
\quad W^{\scriptsize\left\{ \begin{smallmatrix} E \\ H \end{smallmatrix} \right\}}_{m,n}(j), & j = \left\lfloor \kappa_{m,n} \right\rfloor \\
\quad 0, & \text{otherwise}
\end{cases}
\end{equation}
%
%
%
%
and
\begin{multline}
\boldsymbol {\tilde I}^{(j)}=\begin{bmatrix}
\tilde I_j^{(0)} \\
\tilde I_j^{(1)} \\
 \vdots\\
 \tilde I_j^{N}
\end{bmatrix}, \mathbf V_E^{(j)}=\begin{bmatrix}
E^{\rm inc}_z{\rm }(\boldsymbol \rho_1,(j+1)\Delta t)\\
E^{\rm inc}_z{\rm }(\boldsymbol \rho_2,(j+1)\Delta t) \\
 \vdots\\
E^{\rm inc}_z{\rm }(\boldsymbol \rho_M,(j+1)\Delta t)
\\
\end{bmatrix}, \\\mathbf V_H^{(j)}=\begin{bmatrix}
\mathbf {\hat z}\cdot \left(\mathbf {\hat n}(\boldsymbol \rho_1)\times \boldsymbol H^{\rm inc}(\boldsymbol \rho_1,(j+1)\Delta t)  \right)\\
\mathbf {\hat z}\cdot \left(\mathbf {\hat n}(\boldsymbol \rho_2)\times \boldsymbol  H^{\rm inc}(\boldsymbol \rho_2,(j+1)\Delta t)  \right) \\
 \vdots\\
\mathbf {\hat z}\cdot \left(\mathbf {\hat n}(\boldsymbol \rho_M)\times \boldsymbol H^{\rm inc}(\boldsymbol \rho_M,(j+1)\Delta t)  \right)
\\
\end{bmatrix}, ~j=0,1,\ldots
\end{multline}
%
%
%
%
Solving this set of matrix equations is equivalent to performing a sequential solution of a block-lower triangular block-Toeplitz matrix equation, which encapsulates the entire spatio-temporal system and is expressed as follows:
\begin{equation}
\begin{bmatrix}
[\mathbf Z^{(0)}] & \mathbf{0} & \mathbf{0} & \cdots & \mathbf{0} & \cdots & \mathbf{0} \\
[\mathbf Z^{(1)}] & [\mathbf Z^{(0)}] & \mathbf{0} & \cdots & \mathbf{0} &  \cdots & \mathbf{0} \\
[\mathbf Z^{(2)}] & [\mathbf Z^{(1)}] & [\mathbf Z^{(0)}] & \cdots & \mathbf{0} &  \cdots & \mathbf{0} \\
\vdots & \vdots & \vdots & \ddots & \vdots  & \ddots & \vdots \\
[\mathbf Z^{(j)}] & [\mathbf Z^{(j-1)}] & [\mathbf Z^{(j-2)}] & \cdots &[\mathbf Z^{(j-k')}]  & \cdots & \mathbf{0} \\
\vdots & \vdots & \vdots & \vdots & \vdots & \vdots & \vdots &  
\end{bmatrix}\, \begin{bmatrix}
\boldsymbol {\tilde I}^{(0)} \\
\boldsymbol {\tilde I}^{(1)} \\
\boldsymbol {\tilde I}^{(2)}\\
 \vdots\\
 \boldsymbol {\tilde I}^{(k')}\\
 \vdots\\
\end{bmatrix}=\begin{bmatrix}
\mathbf V^{(0)} \\
\mathbf  V^{(1)}\\
\mathbf  V^{(2)}\\
 \vdots\\
\mathbf V^{(k')}\\
\vdots\\
\end{bmatrix}
\label{eq:linear_system}
\end{equation}
where 
%
%
%
\begin{subequations}
\begin{equation}
[\mathbf Z^{(i)}]=[\mathbf Z^{(i)}_E]-\sum_{k'=0}^{i}\zeta_s\Bigl((i-k')\Delta t\Bigr)\Bigl([\mathbf Z^{(k')}_H]-[\mathbf Z^{(k'-1)}_H]\Bigr)
\end{equation}
\begin{equation}
\mathbf V^{(i)}=\mathbf V^{(i)}_E-\sum_{k'=0}^{i}\zeta_s\Bigl ((i-k')\Delta t\Bigr)\Bigl(\mathbf V^{(k')}_H- \mathbf V^{(k'-1)}_H\Bigr )
\end{equation}
\label{eq:matrix_equations}
\end{subequations}%
 The solution of (\ref{eq:linear_system}) leads to the computation of the amplitudes $\tilde I_n^{(k')}$, and subsequently to the calculation of the scattered fields using (\ref{eq:discrete_electric}) and (\ref{eq:discrete_magnetic}).

The generalized matrix equation  (\ref{eq:generalized_2}) can represent either an explicit or an implicit discretization scheme, depending on the spatio-temporal configuration of auxiliary sources and matching points. To attain an explicit discretization scheme, two requirements must be fulfilled. First, the geometry of the scatterer must permit the placement of auxiliary sources such that the source nearest to a given matching point is also strictly closer to it than to any other matching point. Second, the time step $\Delta t$ must be sufficiently small to ensure that, at the end of the first interval, only the nearest auxiliary source contributes to each matching point. This setup results in a square system matrix that can be solved via direct inversion. In contrast, an implicit discretization scheme arises when the time intervals are large enough to allow multiple auxiliary sources to influence each matching point during the same time step. Alternatively, the implicit scheme may also be achieved by choosing the number $M$ of matching points to be larger than the number $N$ of auxiliary sources, resulting in an overdetermined system. Such systems are typically solved using a least-squares approach. 

Although the implicit formulation is computationally more demanding, it often exhibits improved numerical stability and reduced late-time errors compared to its explicit counterpart. In our numerical experiments, the implicit scheme is adopted by selecting the number of matching points to be greater than the number of auxiliary sources.

\section{Numerical Results}
In this section, we present numerical results related to the theoretical analysis discussed in Section~\ref{sec:theory}. A variety of geometries were used, and simulations were performed for both types of polarization, namely $\rm TM_z$ and $\rm TE_z$. Two types of incident waves were also considered: (i) a transient plane wave, and (ii) a transient cylindrical wave generated by a current filament. Additionally, several values of the conductivity were used for the scatterer to investigate the method's robustness under different material properties. These choices ensure the generality of the conclusions regarding the effectiveness of the proposed method.

\subsection{Circular cylinder}
\label{sec:circular}

We begin our numerical study with the case of a circular cylinder of radius \( \rho_c = 0.3~\mathrm{m} \), illuminated by a \(\mathrm{TM}_z\) polarized transient plane wave. The incident electric field is given by
\begin{equation}
\boldsymbol {E}^{\rm inc}(\boldsymbol \rho, t) = \exp\left(-\frac{(4t - \boldsymbol \rho \cdot \mathbf{\hat{x}}/c)^2}{\tau^2}\right) \mathbf{\hat{z}},
\label{eq:incident}
\end{equation}
where \( \tau = \rho_c / c \) defines the temporal width of the pulse, corresponding to the time required for the wave to travel a distance equal to the radius of the cylinder. This choice ensures that the spatial extent of the pulse is approximately equal to the diameter \( 2\rho_c \) of the circular cylinder. The auxiliary surface \( C_{\rm aux} \) is defined as a concentric circle with radius \( \rho_{\rm aux} = 0.75 \rho_c \). The cylinder is non-magnetic with relative dielectric permittivity \( \varepsilon_r = 1 \) and electric conductivity \( \sigma = 15~\mathrm{S/m} \).

The temporal step size \( \Delta t \) is expressed in terms of the Nyquist sampling interval \( \Delta t_0 = 1/(2f_0) \), where \( f_0 \) denotes the effective upper frequency bound of the pulse spectrum. This frequency is selected such that the amplitude in the spectral domain satisfies
\begin{equation}
E^{\mathrm{inc}}_{z}|_{f = f_0} = 10^{-4} \, E^{\mathrm{inc}}_{z}|_{f = 0}.
\end{equation}
Based on this condition, the cutoff frequency and sampling interval are approximated as \( f_0 \approx \frac{1.932}{\tau} \) and \( \Delta t_0 \approx \frac{\tau}{3.864} \), respectively. 
Then, we obtain \( \tau = 10^{-9}~\mathrm{s} \), \( f_0 = 1.932~\mathrm{GHz} \), and \( \Delta t_0 \approx 2.59 \times 10^{-10}~\mathrm{s} \). Any time step \( \Delta t \le \Delta t_0 \) is sufficient to resolve the scattered field. However, as \( \Delta t \) decreases, a correspondingly larger number of auxiliary sources and matching points is required to meet the criteria of the implicit formulation described in Section~\ref{sec:theory}.

To investigate the influence of the time step \( \Delta t \)—and, by extension, the number \( N \)  of auxiliary sources and \( M \) of matching points—three numerical experiments were performed with \( \Delta t = (0.9, 0.4, 0.1)\, \Delta t_0 \). In each case, the number of matching points was selected to be four times the number of auxiliary sources, specifically \( N = 10,~15,~30 \) and \( M = 40,~60,~120 \), respectively. The scattered field was evaluated at three observation points with polar coordinates \( \rho = 3\rho_c \) and \( \phi = 0,~\pi/2,~\pi \), corresponding to the forward, normal, and backward scattering directions with respect to the incident wave. 
The discrete-time values of \( \zeta_s(t) \) used in (\ref{eq:matrix_equations}) were obtained by applying the Inverse Fast Fourier Transform (IFFT) to frequency-domain samples of \( \zeta_s(\omega) \), computed over the frequency range \( f_{\mathrm{min}}=0.0001~\mathrm{GHz} \) to $f_{\mathrm{max}}=1/\Delta t$ with a frequency step $f_{\mathrm{step}}=0.0001~\rm GHz$. This number of frequency-domain samples is much larger than the minimum required by the Nyquist theorem, hence, ensuring accurate time-domain reconstruction.

\begin{figure}[htb!]
    \subfigure[]{\includegraphics[width=0.47\textwidth]{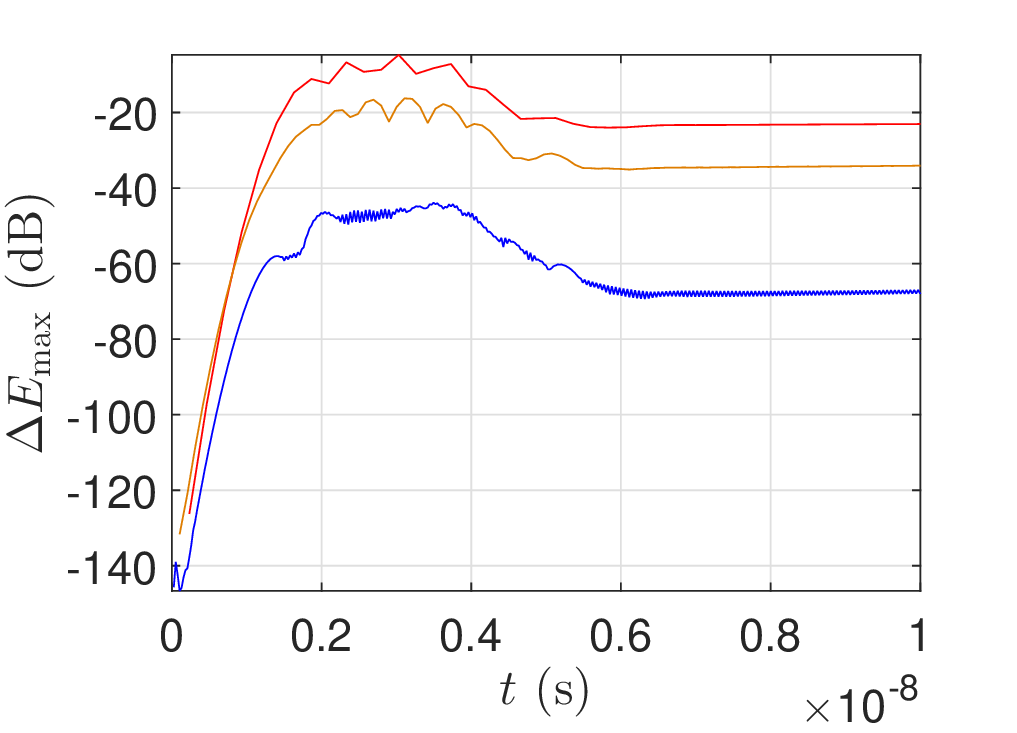}}
    \centering
    \subfigure[]{\includegraphics[width=0.47\textwidth]{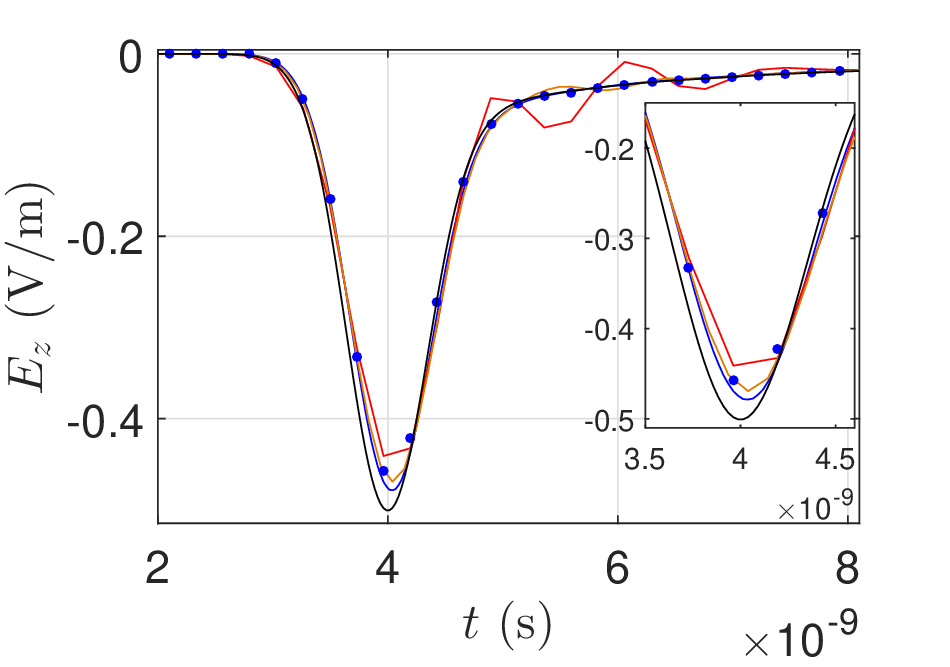}} 
    \subfigure[]{\includegraphics[width=0.47\textwidth]{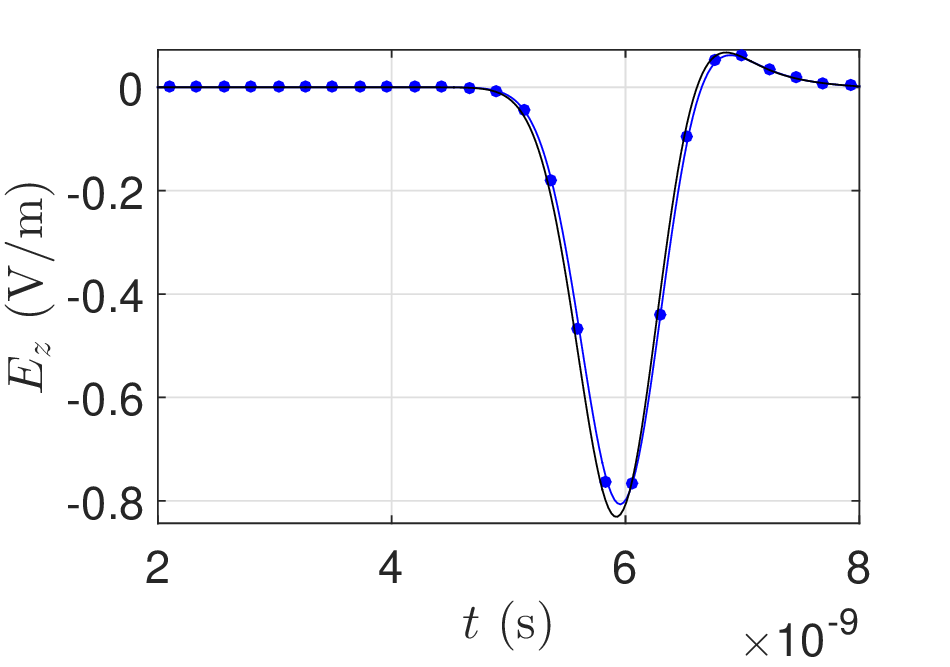}}
    \centering
    \subfigure[]{\includegraphics[width=0.47\textwidth]{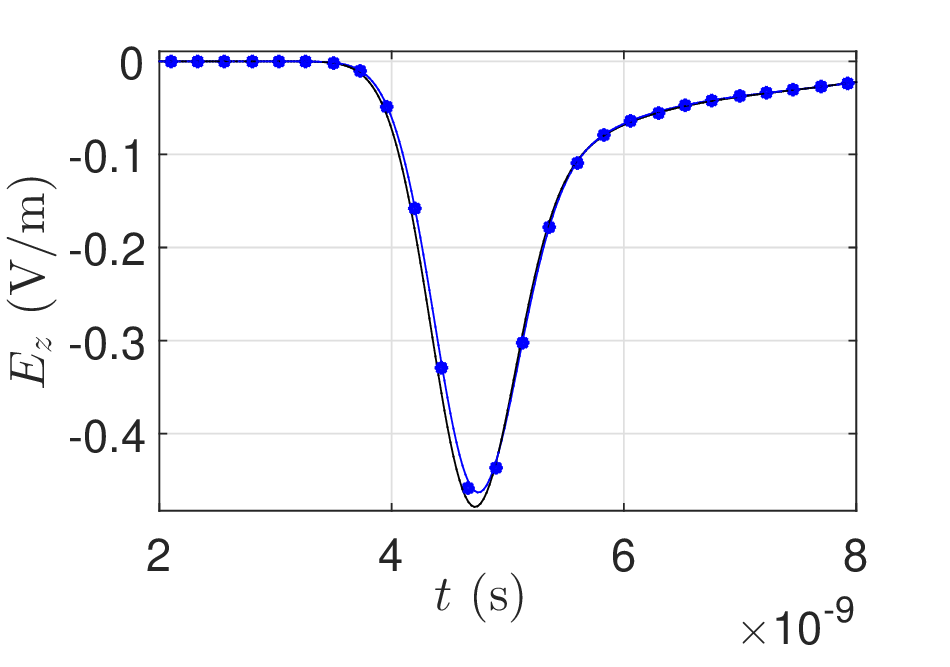}} 
    \caption{Circular cylinder with $\rho_c=0.3~\rm m$ and $\sigma=15~\rm S/m$ under a $\rm TM_z$ transient plane wave incidence: (a) boundary condition error $\Delta E_{\rm max}$ (b) amplitude of the scattered electric field $E_z$ at the observation point $(-0.9~\rm m, 0)$, (c) $E_z$ at $(0, 0.9~\rm m)$, and (d) $E_z$ at $(0.9~\rm m, 0)$. 
    For (b)--(d) blue dots correspond to analytic solver results and blue solid lines correspond to MAS results for $N=30$, $M=120$, and $\Delta t=0.1\Delta t_0$ as is for (b).  For (b)--(c) black solid line corresponds to the amplitude of the back scattered field for a PEC scatterer as obtained by an analytic solver. For (a) and (b), red solid line corresponds to  $N=10$, $M=40$, and $\Delta t=0.9\Delta t_0$ and orange solid line for $N=15$, $M=60$, and $\Delta t=0.4\Delta t_0$.}
    \label{fig:circle_2}
\end{figure}

To provide a quantitative measure of the method's numerical performance, we begin by defining and evaluating the \textit{boundary condition error}. Although the accuracy of the proposed MAS approach is subsequently validated against an analytical solution, analyzing the boundary condition error offers independent insight into how the spatiotemporal resolution---governed by \( \Delta t \), \( N \), and \( M \)---influences the solution quality. The boundary condition error is defined as
\begin{multline}
\Delta E_{\rm max}~(\mathrm {dB})= 20\log_{10}\\\left[  \frac{\underset{\boldsymbol{\rho} \in C}{\max} \left|E^{\rm inc}_z(\boldsymbol{\rho},t) + E^{\mathrm{sc}}_z(\boldsymbol{\rho},t) - \zeta_s(t)*\left[ \mathbf {\hat n}\times \frac{\partial }{\partial t}\left (\boldsymbol H^{\rm inc}(\boldsymbol{\rho},t) + \boldsymbol H^{\mathrm{sc}}(\boldsymbol{\rho},t) \right) \right]  \right|}{\underset{t}{\max}\left( \frac{1}{2}\,\underset{\boldsymbol{\rho} \in C}{\max} \Biggl | E^{\rm inc}_z(\boldsymbol{\rho},t) + \zeta_s(t)*\left[ \mathbf {\hat n}\times \frac{\partial }{\partial t}\boldsymbol H^{\rm inc}(\boldsymbol{\rho},t) \right]\Biggr| \right) }\right]
\end{multline}
\noindent Figure~\ref{fig:circle_2}(a) illustrates the boundary error as a function of time, clearly showing a significant decrease as the spatiotemporal resolution increases. This trend reinforces the expectations regarding parameter sensitivity and supports the subsequent validation against analytical results.

To validate the results of the MAS, we employed an analytical frequency-domain solver to generate a dense set of frequency-domain data at the three observation points within the frequency range \( 0.0001~\rm GHz \le f \le 2.145~GHz \). An IFFT was then applied to these samples to obtain the corresponding time-domain results. The MAS and analytical results are presented in Figs.~\ref{fig:circle_2}(b), (c), and (d), where the amplitude of the scattered electric field \( E_z \) is plotted with respect to time \( t \).

In the back-scattered regime (Fig.~\ref{fig:circle_2}(b)), results are presented for all three parameter sets \( (\Delta t, N, M) \), in order to clearly illustrate their influence on the accuracy of the solution. The same figure also includes the plot of the scattered electric field for the case of a Perfect Electric Conductor (PEC) scatterer (as computed by the analytic solver), in order to highlight the difference compared to the case of a conductor with finite conductivity. As expected, the most accurate MAS results are obtained for \( \Delta t = 0.1\, \Delta t_0 \), \( N = 30 \), and \( M = 120 \) (as observed in the inset plot of Fig.~\ref{fig:circle_2}(b)), which are also the parameters used to generate the plots in Figs.~\ref{fig:circle_2}(c) and (d). Overall, the derived MAS results are characterized by high accuracy, as further supported by the boundary condition analysis.


%

\begin{figure}[htb!]
\centering
    \subfigure[]{\includegraphics[width=0.47\textwidth]{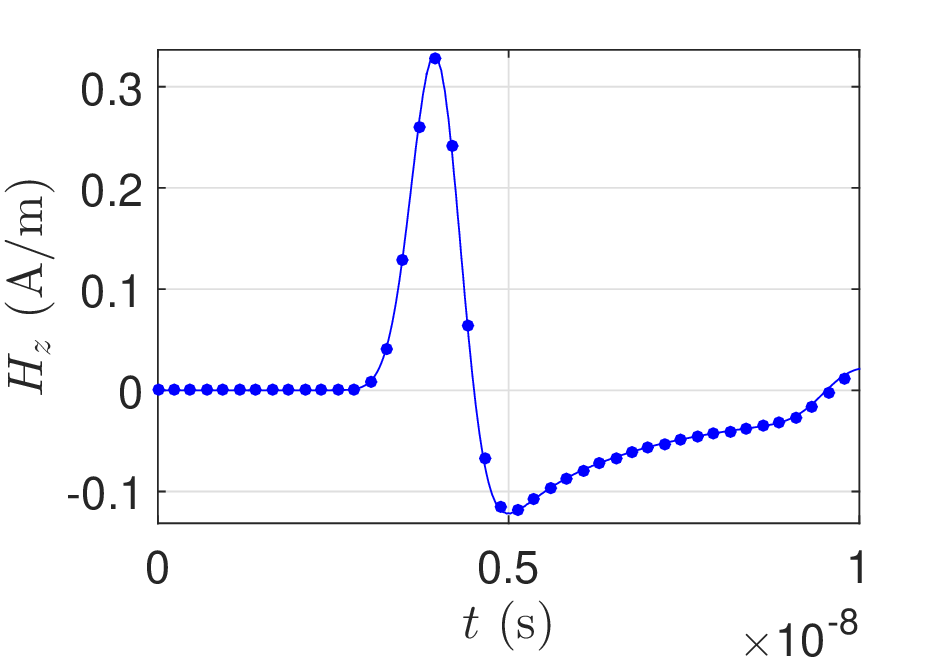}}
    \centering
    \subfigure[]{\includegraphics[width=0.47\textwidth]{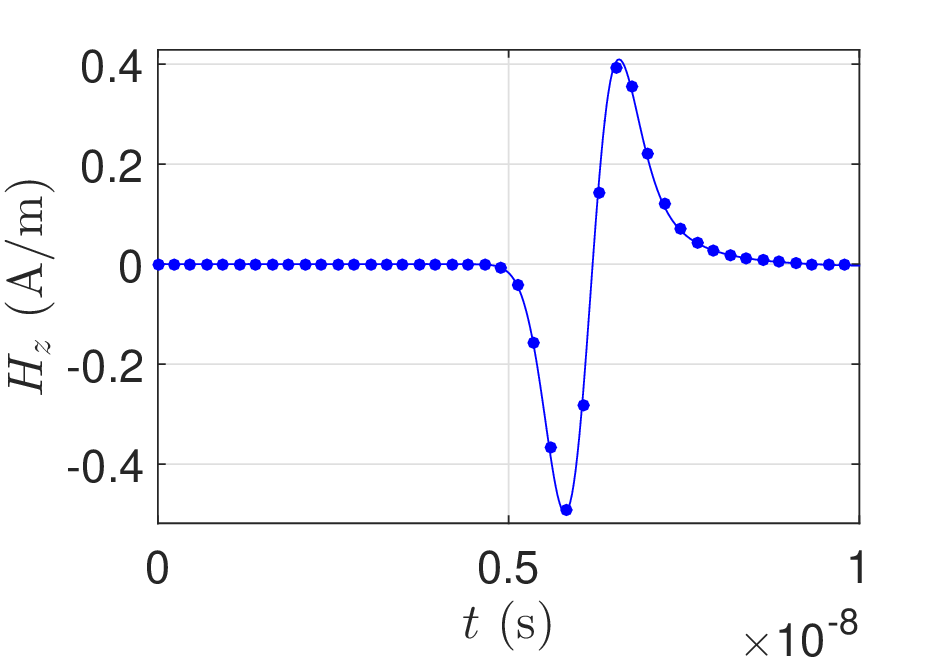}} 
    \subfigure[]{\includegraphics[width=0.47\textwidth]{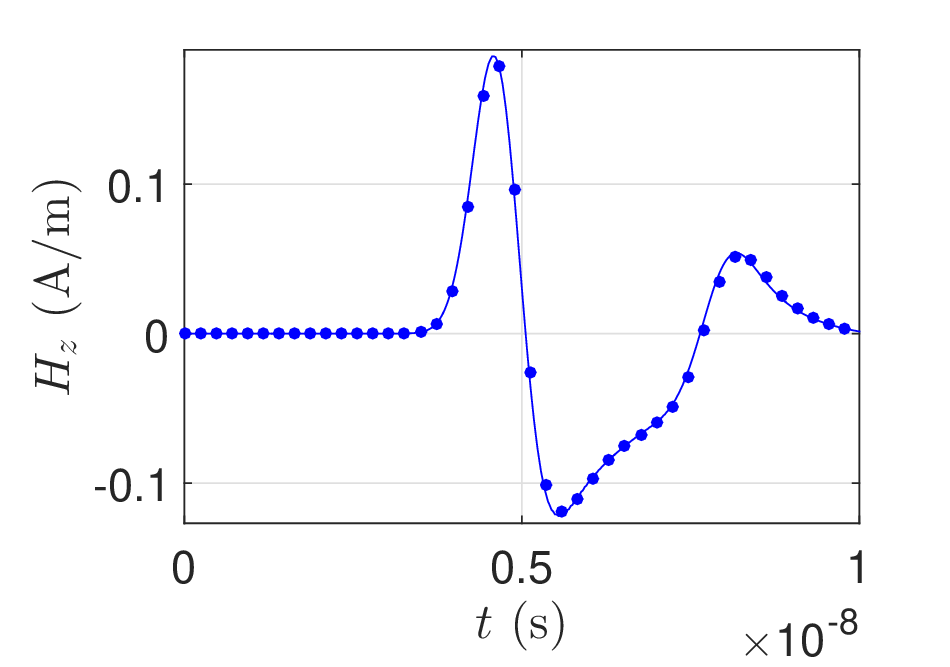}}
    \caption{Circular cylinder with $\rho_c=0.3~\rm m$ and $\sigma=15~\rm S/m$ under a $\rm TE_z$ transient plane wave incidence: (a) amplitude of the scattered magnetic field $H_z$ at the observation point $(-0.9~\rm m, 0)$, (b) $H_z$ at $(0, 0.9~\rm m)$, (c) $H_z$ at $(0.9~\rm m, 0)$.  For all plots, blue dots correspond to the analytic solver results and blue solid lines correspond to MAS results for $N=30$, $M=120$, and $\Delta t=0.1 \Delta t_0$.}
    \label{fig:circular_TE}
\end{figure}

The case of a $\rm TE_z$ incident wave, where its magnetic field \( \boldsymbol H^{inc}(\boldsymbol \rho,t)\) is described by the right hand side of (\ref{eq:incident}), is treated as mentioned earlier via interchanging the magnetic and electric sources. The respective results, computed with high spatiotemporal resolution, are shown in Figs.~\ref{fig:circular_TE} (a), (b), and (c) as plots of the amplitude of the scattered magnetic field at three observation points.
\begin{figure}[htb!]
\centering
    \subfigure[]{\includegraphics[width=0.47\textwidth]{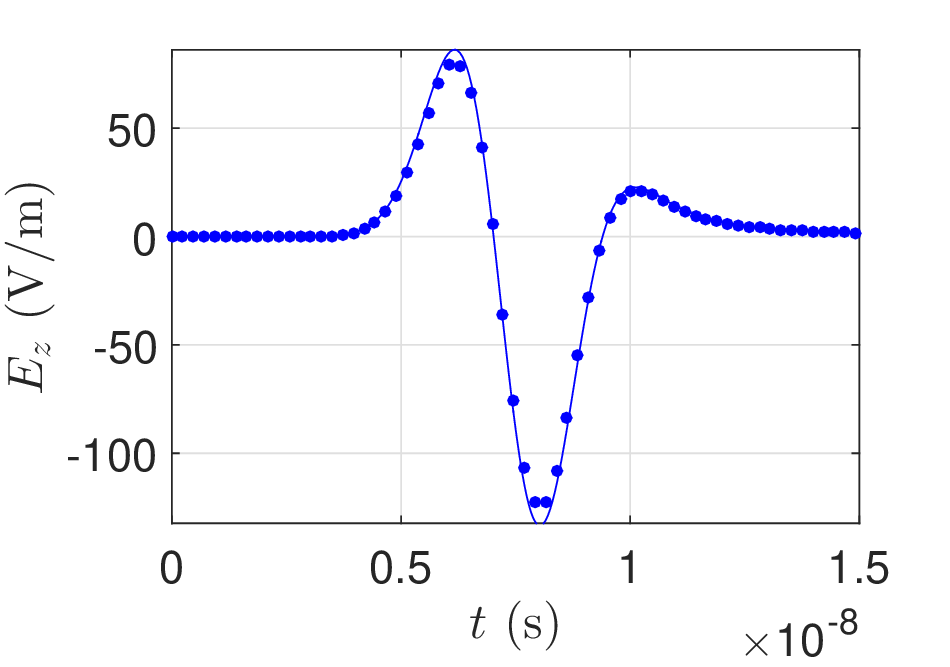}} 
    \subfigure[]{\includegraphics[width=0.47\textwidth]{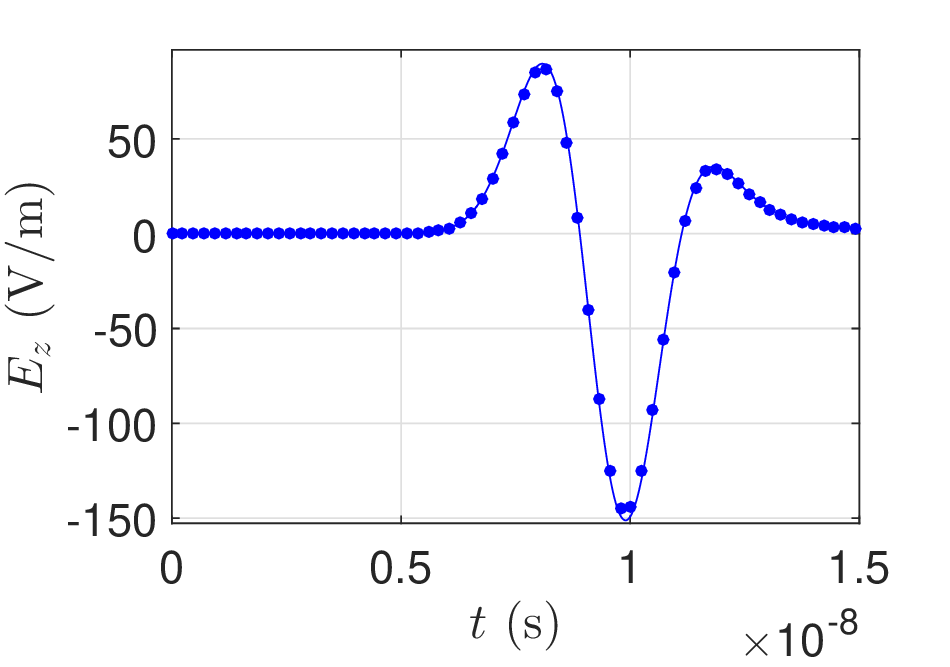}}
    \centering
    \subfigure[]{\includegraphics[width=0.47\textwidth]{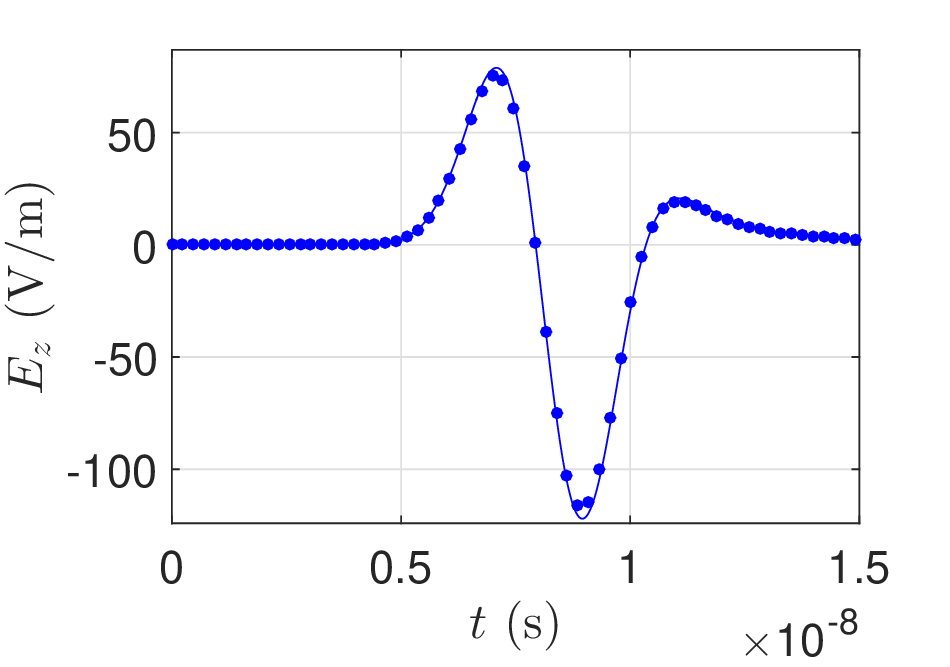}} 
    \caption{Circular cylinder with $\rho_c=0.3~\rm m$ and $\sigma=5~\rm S/m$ under a $\rm TM_z$ transient cylindrical wave of an electric line source $I(t)$ positioned at $(x_s, y_s)=(0.6~\rm m,0)$: (d) amplitude of the scattered electric field $E_z$ at the observation point $(-0.9~\rm m, 0)$, (e) $E_z$ at $(0, 0.9~\rm m)$, (f) $E_z$ at $(0.9~\rm m, 0)$. 
    For all plots, blue dots correspond to the analytic solver results and blue solid lines correspond to MAS results for $N=30$, $M=120$, and $\Delta t=0.1 \Delta t_0$.}
    \label{fig:circular_line}
\end{figure}

To further broaden the scope of our investigation and reinforce the generality of the proposed method, we also examine the case of a transient cylindrical wave impinging on a circular cylinder. 
The \( \mathrm{TM}_z \) cylindrical wave is generated by an electric current filament excited by a differentiated Gaussian current pulse, with its temporal profile given by
\begin{equation}
I_0(t) = -\left( \frac{t}{\tau} \right) \exp\left[ 0.5 - 0.5\left( \frac{t}{\tau} \right)^2 \right].
\label{eq:current_pulse}
\end{equation}
The electric and magnetic fields radiated by this source are calculated by (\ref{eq:discrete_electric}) and (\ref{eq:discrete_magnetic}) applied for a single source, i.e., $N=1$. For this scenario, we selected a conductivity of \( \sigma = 5~\mathrm{S/m} \) and set the pulse width parameter to \( \tau = \rho_c/c = 10^{-9}~\mathrm{s} \). Consequently, both the Nyquist sampling interval \( \Delta t_0 \) and the effective upper frequency bound \( f_0 \) retain the same values as those used in the transient plane wave case previously discussed. The amplitude of the scattered electric field \( E_z \) is plotted in Figs.~\ref{fig:circular_line}(a), (b), and (c), corresponding to a high spatio-temporal resolution. The comparison with the analytical solution demonstrates excellent agreement, further validating the accuracy of the proposed method.

\subsection{Elliptical cylinder}
\label{sec:elliptical}
\begin{figure}[htb!]
    \centering
    \subfigure[]{\includegraphics[width=0.47\textwidth]{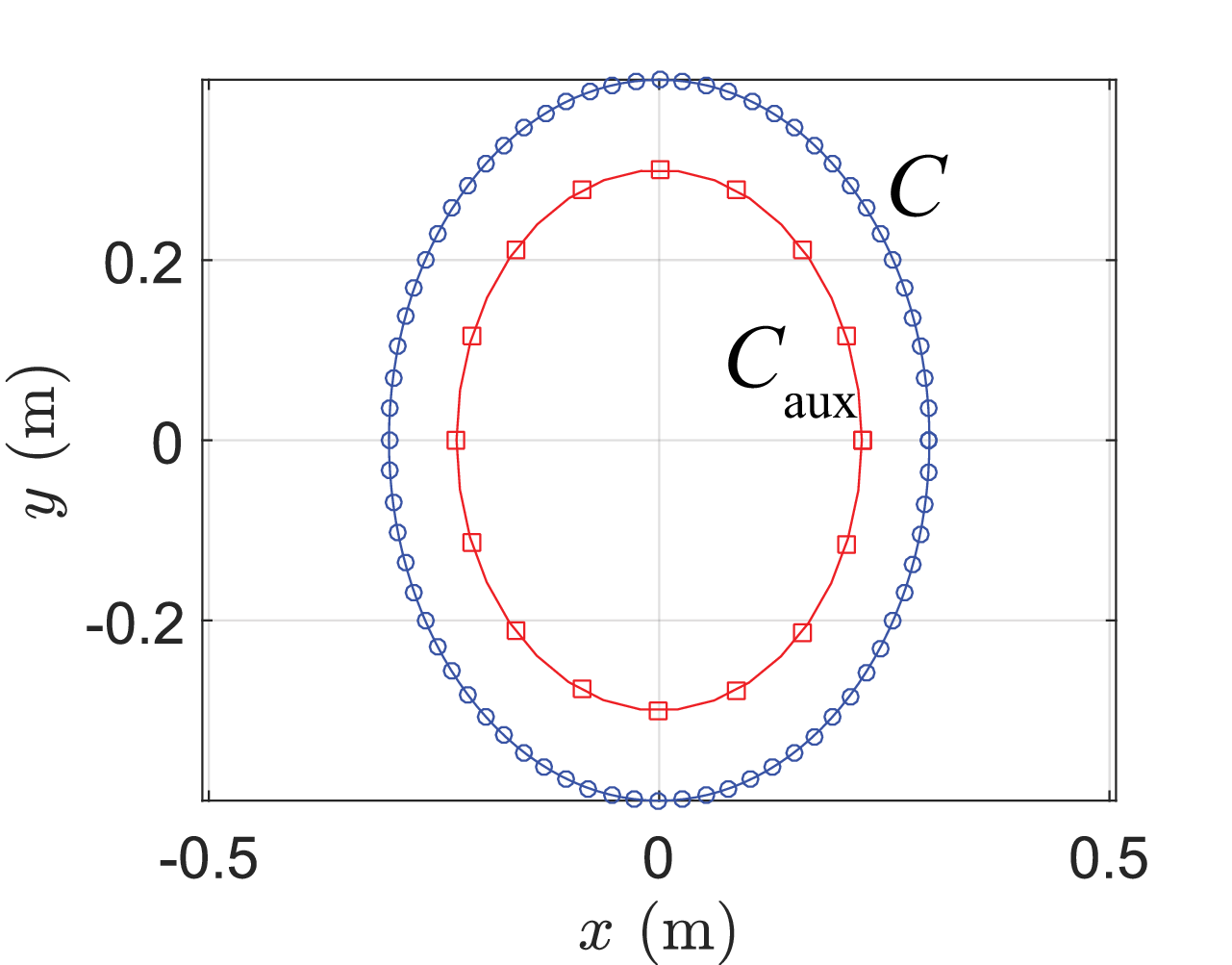}} 
    \subfigure[]{\includegraphics[width=0.47\textwidth]{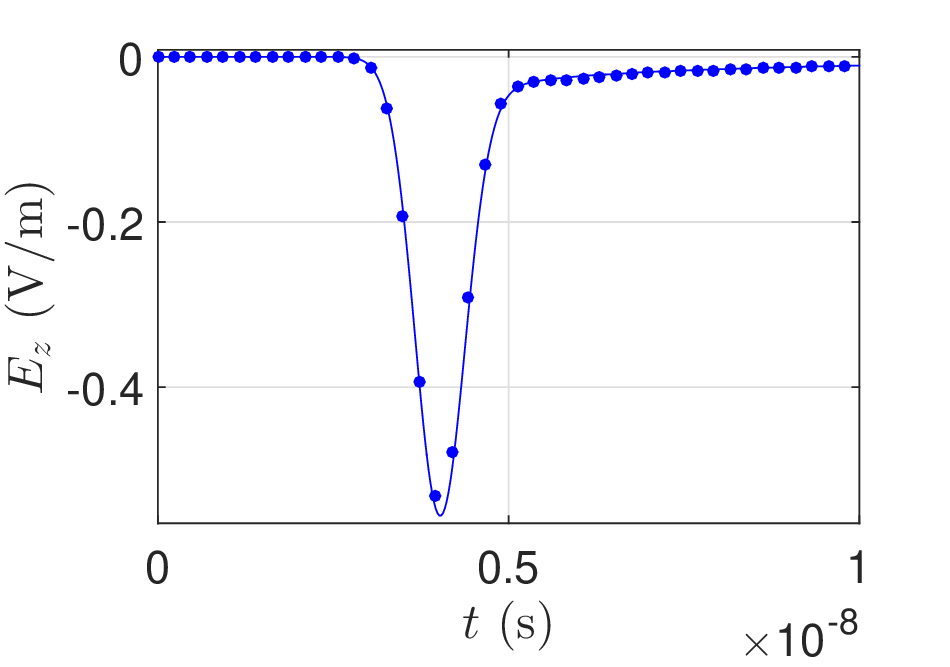}}
    \centering
    \subfigure[]{\includegraphics[width=0.47\textwidth]{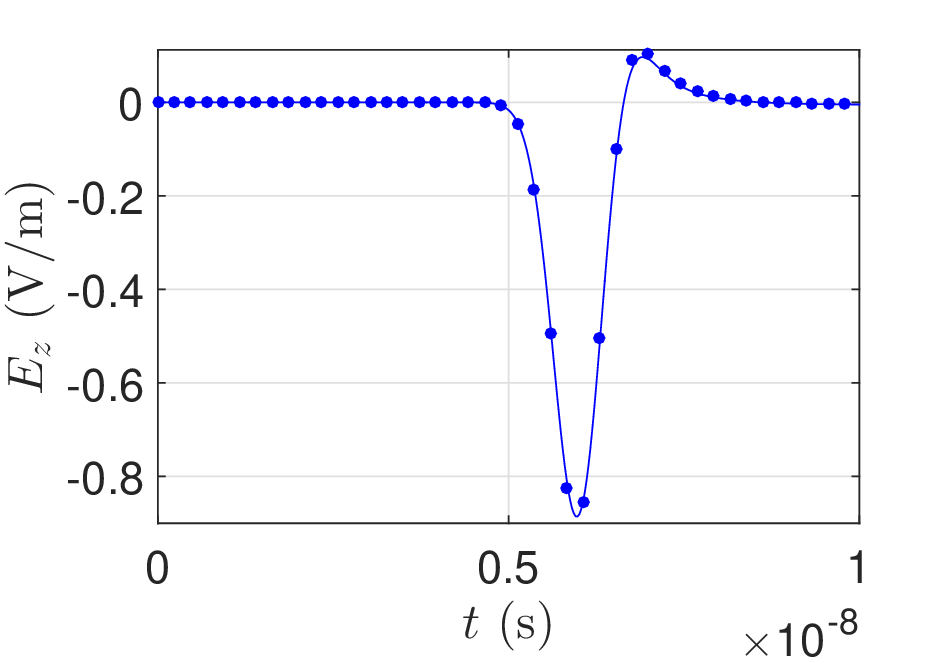}} 
    \subfigure[]{\includegraphics[width=0.47\textwidth]{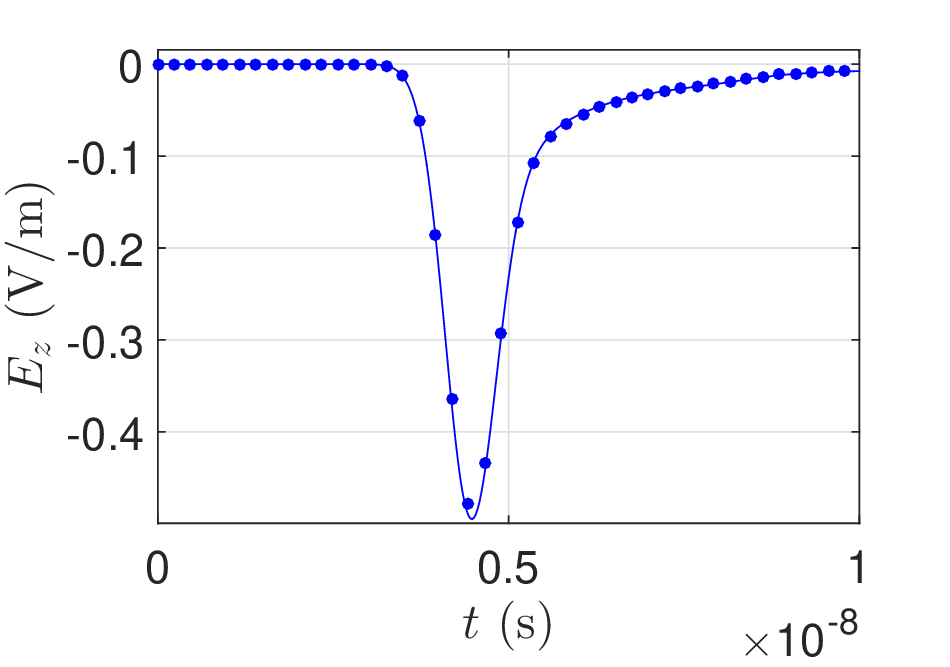}}
    \caption{Elliptical cylinder with $\sigma=10~\rm S/m$ under a $\rm TM_z$ transient plane wave incidence: (a) geometry of the problem, (b) amplitude of the scattered  electric field $E_z$ at the observation point $(-0.9~\rm m, 0)$ (c) $E_z$ at $(0, 0.9~\rm m)$ (d) $E_z$ at $(0.9~\rm m, 0)$. For (a) red squares correspond to auxiliary sources, while for (b), (c), and (d) blue dots correspond to COMSOL results and blue solid lines to MAS results.}
    \label{fig:ellipse}
\end{figure}

The next geometry to be studied is that of the elliptical cylinder with $\sigma=10~\rm S/m$. The semi-major axis is $0.4~\rm m$, while the semi-minor axis is $0.3~\rm m$, as shown in Fig. \ref{fig:ellipse}(a). The same $\rm TM_z$ transient plane wave, as the one considered in Section \ref{sec:circular}, is incident on the cylinder. The auxiliary surface \( C_{\rm aux} \) is a down-scaled version of \( C \) with a scaling factor \( a_{\rm aux} = 0.75 \). We employ a high spatio-temporal resolution with \( N = 34 \) auxiliary sources and \( M = 132 \) matching points, and a time step \( \Delta t = 0.1 \Delta t_0 \).

Validation of the MAS results was carried out using 2145 FD samples  computed with COMSOL Multiphysics over the frequency range \( 0.0001~\mathrm{GHz} \le f \le 2.145~\mathrm{GHz} \), followed by the application of the IFFT. The amplitude of the scattered electric field is plotted at three observation points in Figs.~\ref{fig:ellipse}(b), (c), and (d), where a strong agreement between the MAS and COMSOL results is clearly observed.

\subsection{Super-circular cylinder}
\begin{figure}[htb!]
    \centering
    \subfigure[]{\includegraphics[width=0.47\textwidth]{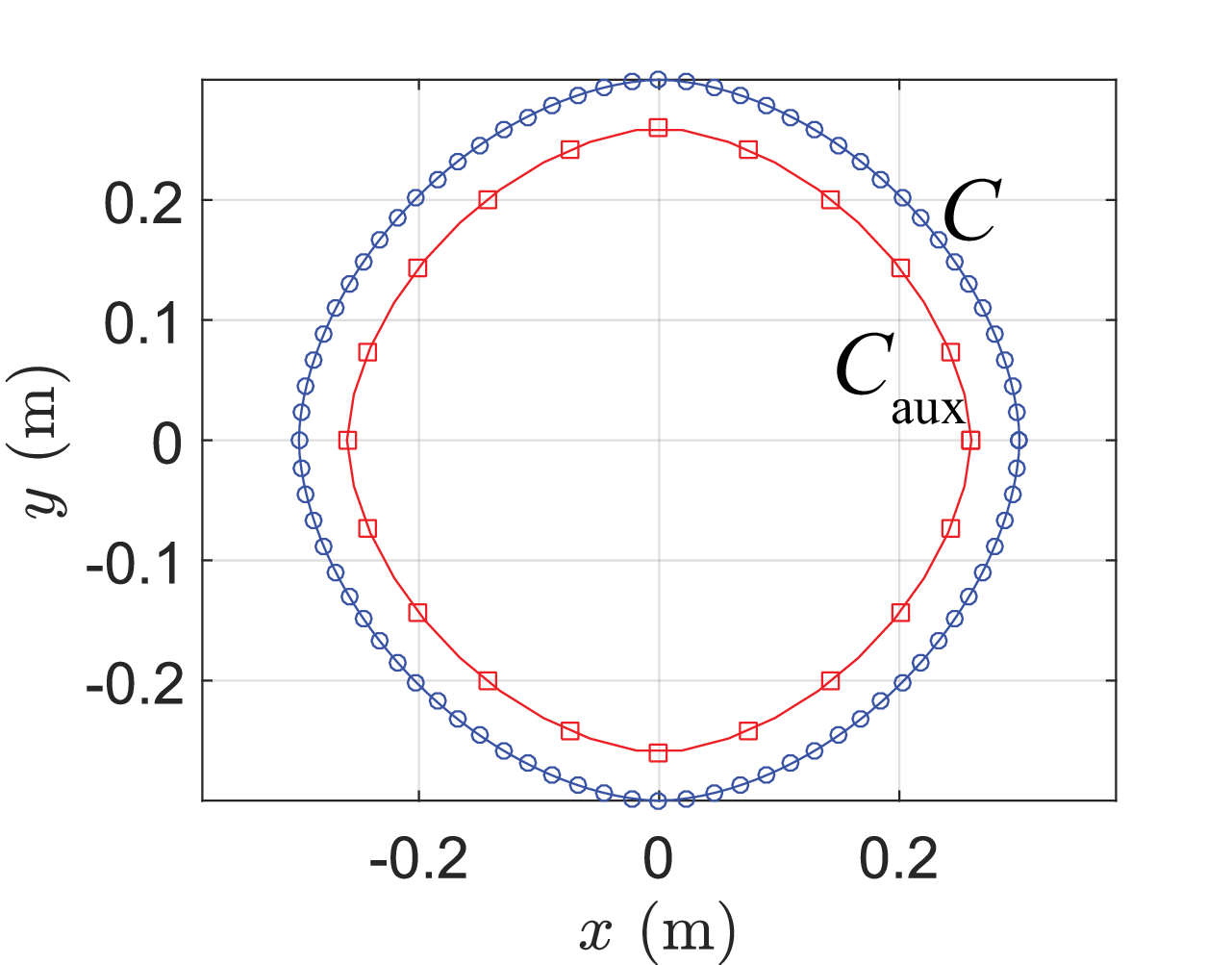}} 
    \subfigure[]{\includegraphics[width=0.47\textwidth]{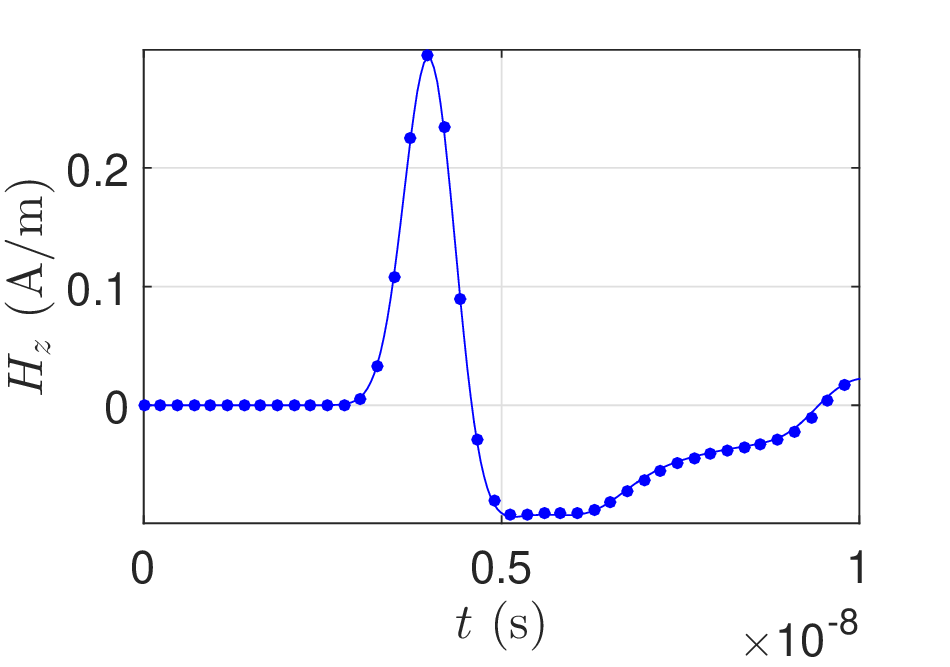}}
    \centering
    \subfigure[]{\includegraphics[width=0.47\textwidth]{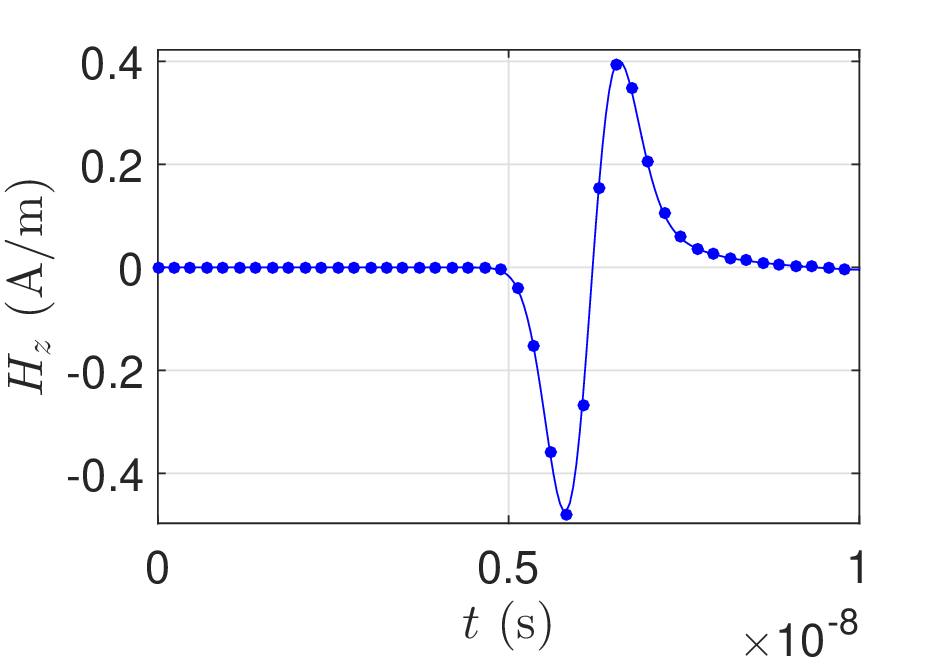}} 
    \subfigure[]{\includegraphics[width=0.47\textwidth]{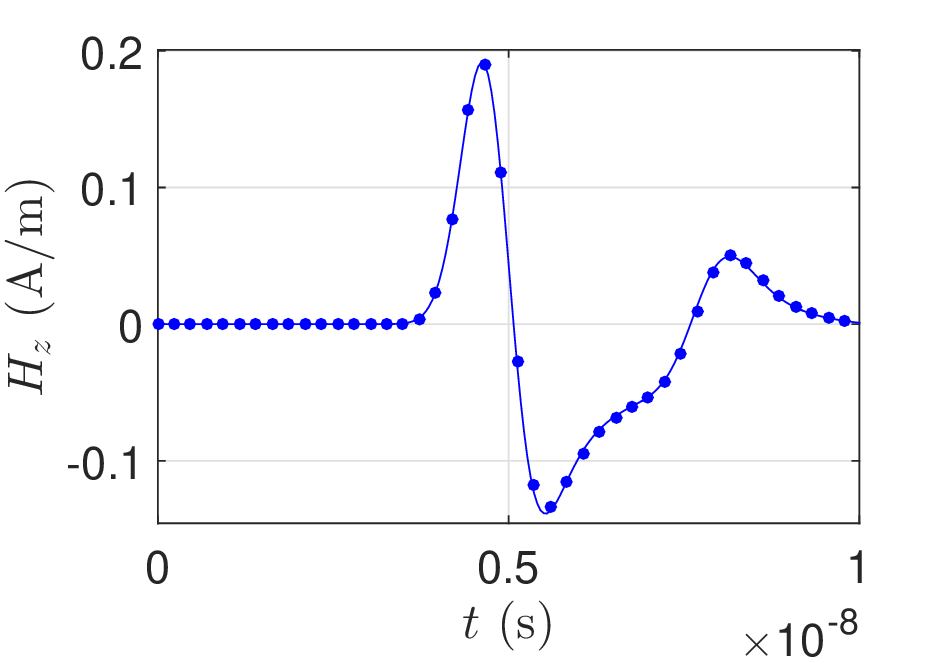}}
    \caption{Super-circular cylinder with $\sigma=15~\rm S/m$ under a $\rm TE_z$ transient plane wave incidence: (a) geometry of the problem, (b) amplitude of the magnetic field $H_z$ at the observation point $(-0.9~\rm m, 0)$, (c) $H_z$ at $(0, 0.9~\rm m)$, (d) $H_z$ at $(0.9~\rm m, 0)$. For (a) red squares correspond to auxiliary sources, while for (b), (c), and (d) blue dots correspond to COMSOL results and blue solid lines to MAS results.}
    \label{fig:super}
\end{figure}

The third geometry under investigation is that of the super-circle with $\sigma=15~\rm S/m$, which is shown in Fig.~\ref{fig:super}(a), and defined by the following parametric equations (in meters),
\begin{subequations}
\begin{equation}
x=0.3\,\mathrm {sgn}(\cos\phi)\left| \cos\phi \right|^{2/n}
\end{equation}
\begin{equation}
y=0.3\,\mathrm {sgn}(\sin\phi)\left| \sin\phi \right|^{2/n}
\end{equation}
\end{subequations}

\noindent with $n=3.5/2$.

A different strategy was employed for the placement of auxiliary sources and matching points compared to the elliptical geometry. Specifically, the matching points were distributed equidistantly along the contour \( C \) (as opposed to using uniformly spaced angles). Subsequently, every fifth matching point was shifted inward by a distance of \( d = 0.04~\mathrm{m} \) along the direction of the normal vectors \( \mathbf{n}_m \) defined at those points. This procedure determined the positions of the auxiliary sources and effectively defined \( C_{\rm aux} \) as an inward offset version of \( C \) separated by a distance \( d \). 

For this geometry, we employed \( N = 34 \) auxiliary sources and \( M = 136 \) matching points, with a time step \( \Delta t = 0.18 \, \Delta t_0 \). The incident field was a \( \mathrm{TE}_z \) transient plane wave, defined as in Section~\ref{sec:circular}. The results, shown in Figs.~\ref{fig:super}(b), (c), and (d), display the amplitude \( H_z \) of the scattered magnetic field. For validation, the same procedure used in the elliptical cylinder was applied, yielding excellent agreement.

\subsection{Rounded-triangular cylinder}
\begin{figure}[htb!]
    \centering
    \subfigure[]{\includegraphics[width=0.47\textwidth]{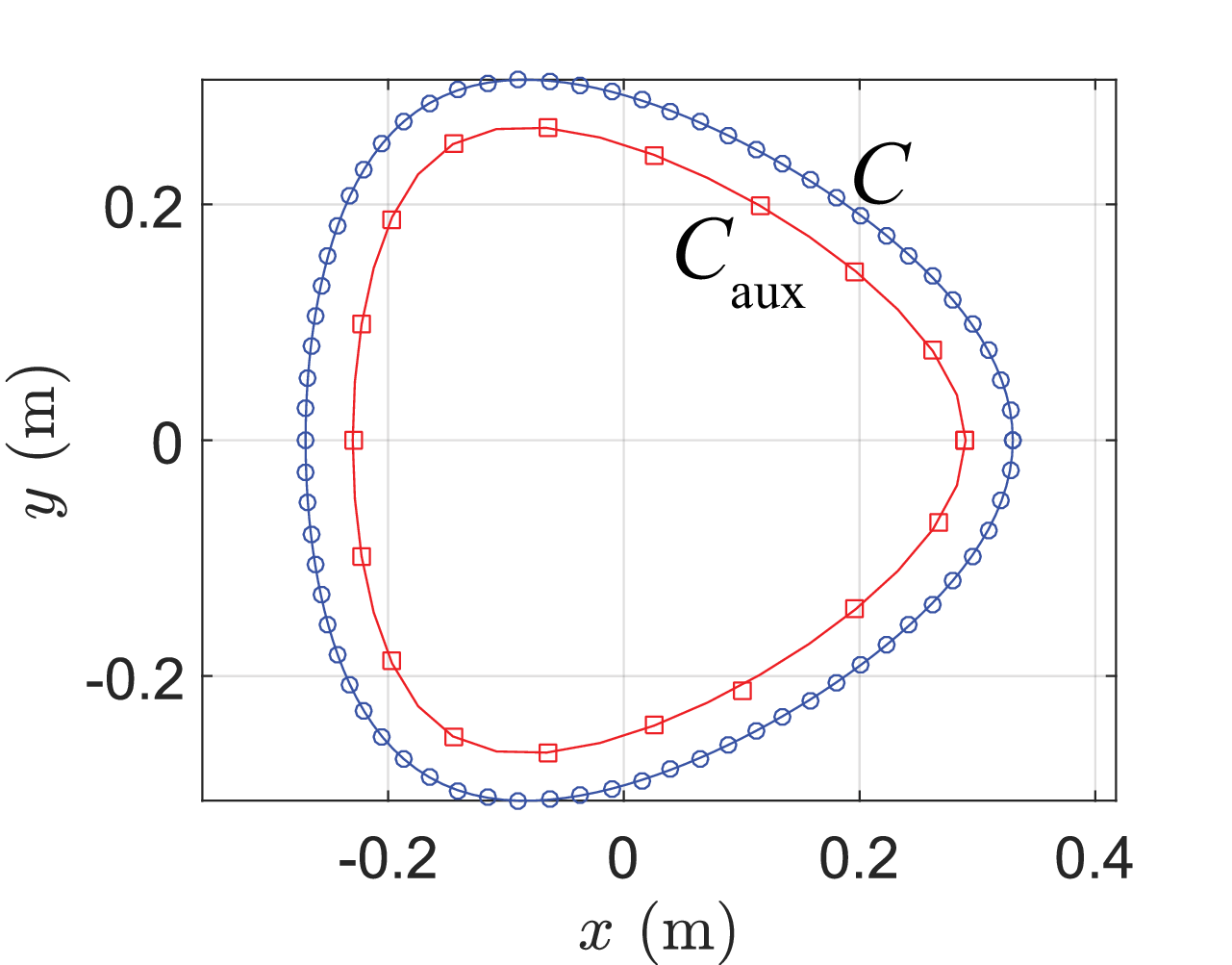}} 
    \subfigure[]{\includegraphics[width=0.47\textwidth]{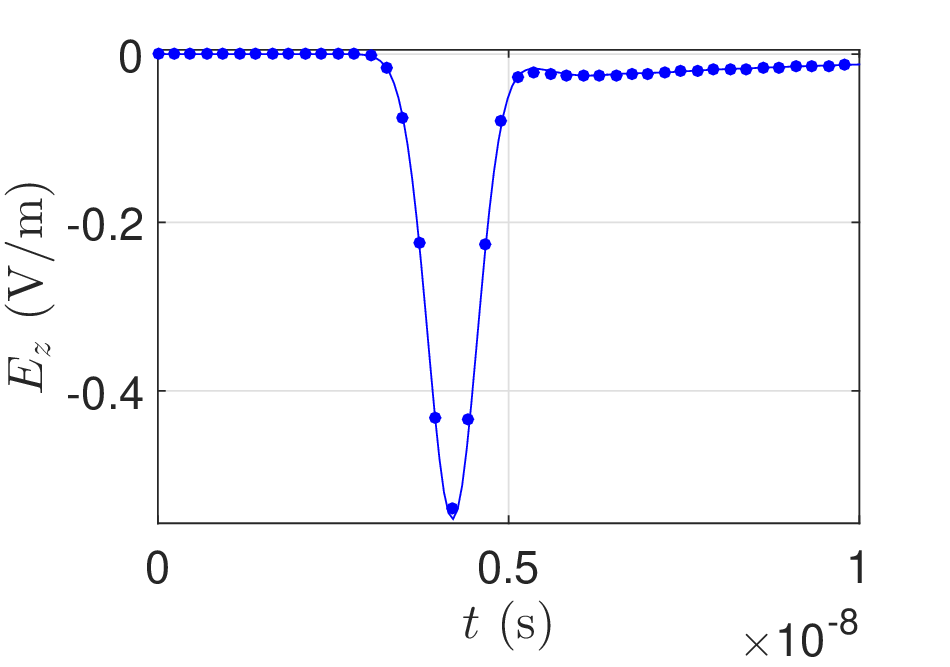}}
    \centering
    \subfigure[]{\includegraphics[width=0.47\textwidth]{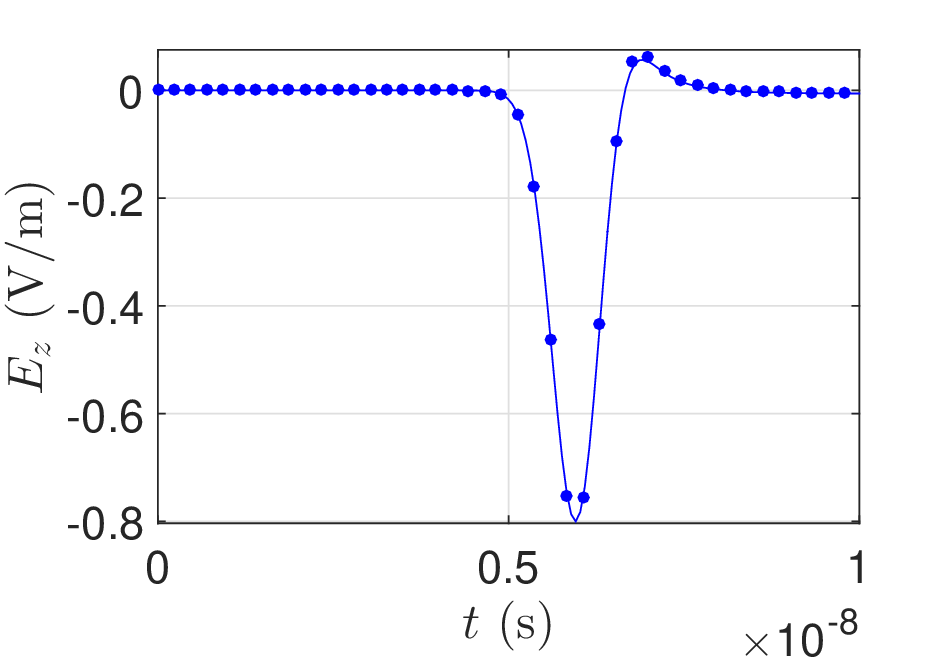}} 
    \subfigure[]{\includegraphics[width=0.47\textwidth]{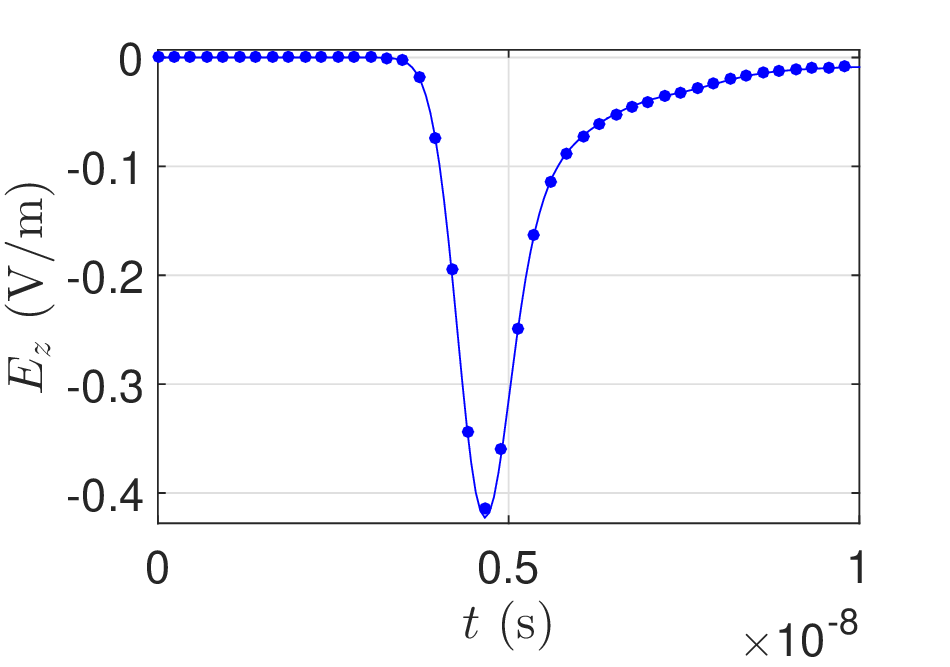}}
    \caption{Rounded-triangular cylinder with $\sigma=15~\rm S/m$ under a $\rm TM_z$ transient plane wave incidence: (a) geometry of the problem, (b) amplitude of the scattered electric field $E_z$ at the observation point $(-0.9~\rm m, 0)$, (c) $E_z$ at $(0, 0.9~\rm m)$, (d) $E_z$ at $(0.9~\rm m, 0)$. For (a) red squares correspond to auxiliary sources and blue circles to matching points, while for (b), (c), and (d) blue dots correspond to COMSOL results and blue solid lines to MAS results.}
    \label{fig:rounded}
\end{figure}

Another interesting geometry is that of the rounded triangle illustrated in Fig. \ref{fig:rounded}(a) and defined by the following equations (in meters)
\begin{subequations}
\begin{equation}
x = 0.3(\cos\phi + 0.1 \cos 2\phi)
\end{equation}
\begin{equation}
y= 0.3(\sin \phi - 0.1 \sin 2\phi)
\end{equation}
\end{subequations}

In this experiment $\sigma=15~\rm{S/m}$, \( N = 32 \), \( M = 128 \), and $\Delta t=0.23 \Delta t_0$, while the \( \mathrm{TM}_z \) incident transient plane of (\ref{eq:incident}) wave was used. 
The same methodology as in the super-circular geometry was applied for the placement of the auxiliary sources and matching points, with \( d = 0.04~\mathrm{m} \). The amplitudes of the scattered electric field $E_z$, as obtained by MAS and COMSOL, are plotted at three observation points in Figs.~\ref{fig:rounded}(b), (c), and (d). 

\subsection{Inverted-elliptical cylinder}
\begin{figure}[htb!]
    \centering
    \subfigure[]{\includegraphics[width=0.47\textwidth]{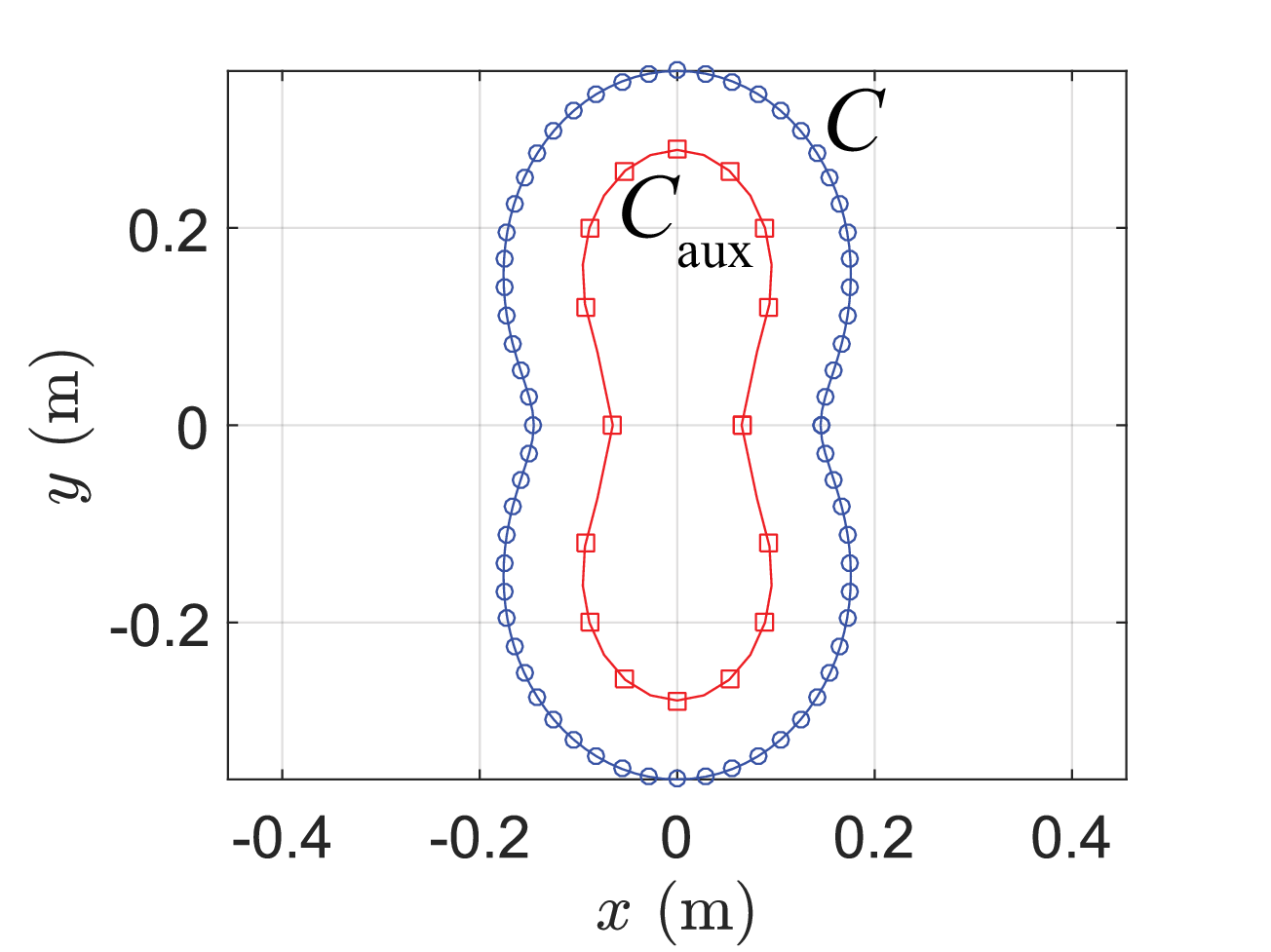}} 
    \subfigure[]{\includegraphics[width=0.47\textwidth]{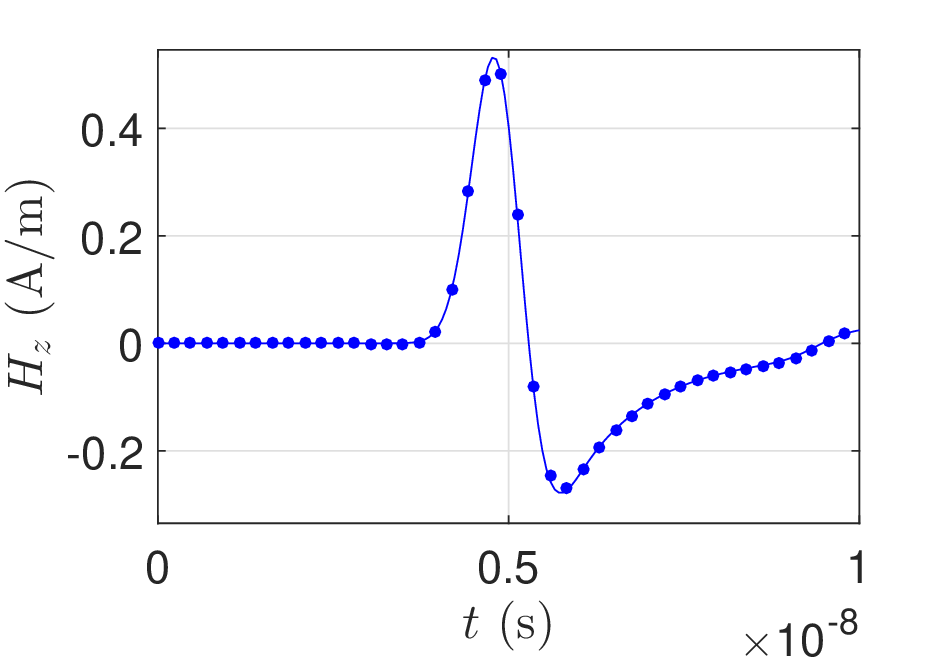}}
    \centering
    \subfigure[]{\includegraphics[width=0.47\textwidth]{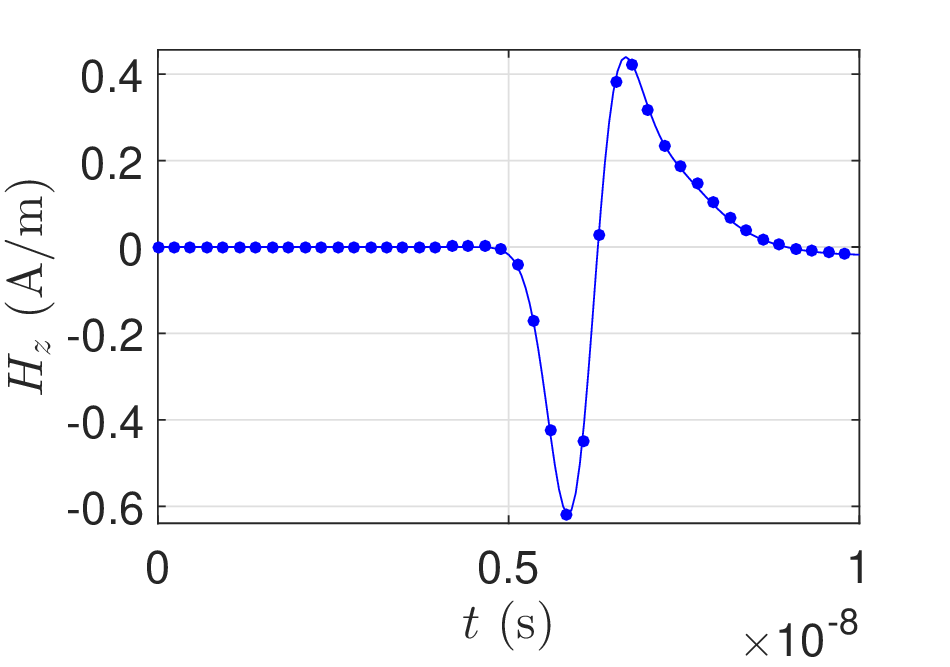}} 
    \subfigure[]{\includegraphics[width=0.47\textwidth]{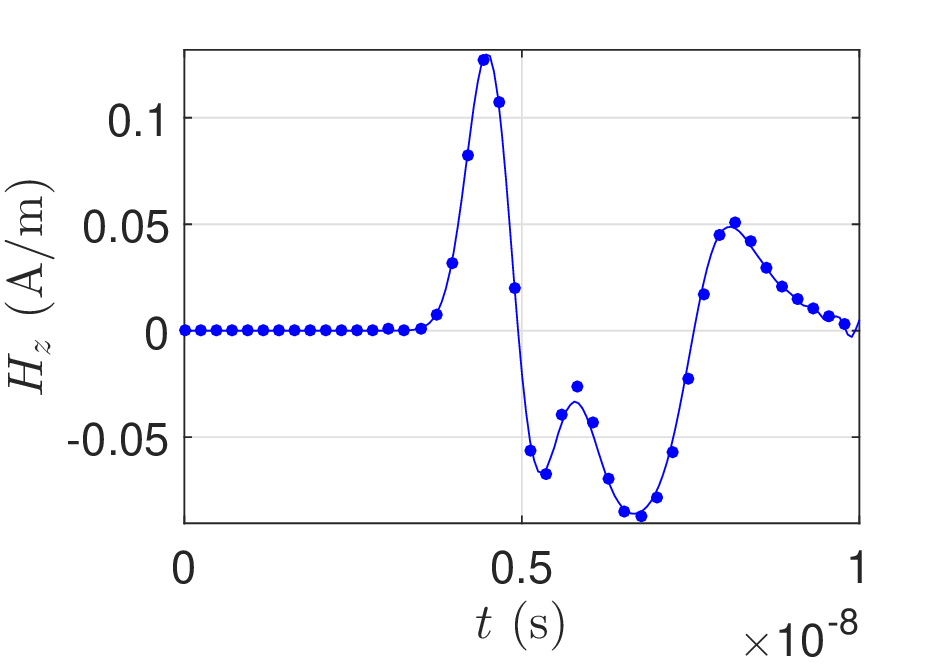}}
    \caption{Inverted-elliptical cylinder with $\sigma=11~\rm S/m$ under a $\rm TE_z$ transient plane wave incidence: (a) geometry of the problem, (b) amplitude of the scattered magnetic field $H_z$ at the observation point $(-0.9~\rm m, 0)$, (c) $H_z$ at $(0, 0.9~\rm m)$, (d) $H_z$ at $(0.9~\rm m, 0)$. For (a) red squares correspond to auxiliary sources and blue circles on matching points while for (b), (c), and (d) blue dots correspond to COMSOL results and blue solid lines to MAS results.}
    \label{fig:bicircle}
\end{figure}

Up to this point, we have treated convex geometries. It is also important to study non-convex cylinders, such as the inverted ellipse shown in Fig.~\ref{fig:bicircle}(a). This geometry is described by the following parametric equations (in meters)
\begin{subequations}
\begin{equation}
x=0.26\frac{\cos\phi}{1+0.7\cos2\phi+0.09}
\end{equation}
\begin{equation}
y=0.14\frac{\sin\phi}{1+0.7\cos2\phi+0.09}
\end{equation}
\end{subequations}

The inverted-elliptical cylinder was analyzed using a conductivity of $\sigma = 11~\rm{S/m}$ and a $\mathrm{TE}_z$ incident transient plane wave. The simulation parameters were set to $N = 32$, $M = 128$ and $\Delta t = 0.23 \Delta t_0$. The auxiliary sources and matching points were placed following the same methodology used for the super-circular geometry, with a spacing of $d = 0.04~\mathrm{m}$. The amplitudes of the scattered electric field $H_z$, computed using both MAS and COMSOL, are shown at three observation points in Figs.~\ref{fig:bicircle}(b), (c), and (d).

\subsection{Planar Geometry}
\begin{figure}[htb!]
    \centering
    \subfigure[]{\includegraphics[width=0.48\textwidth]{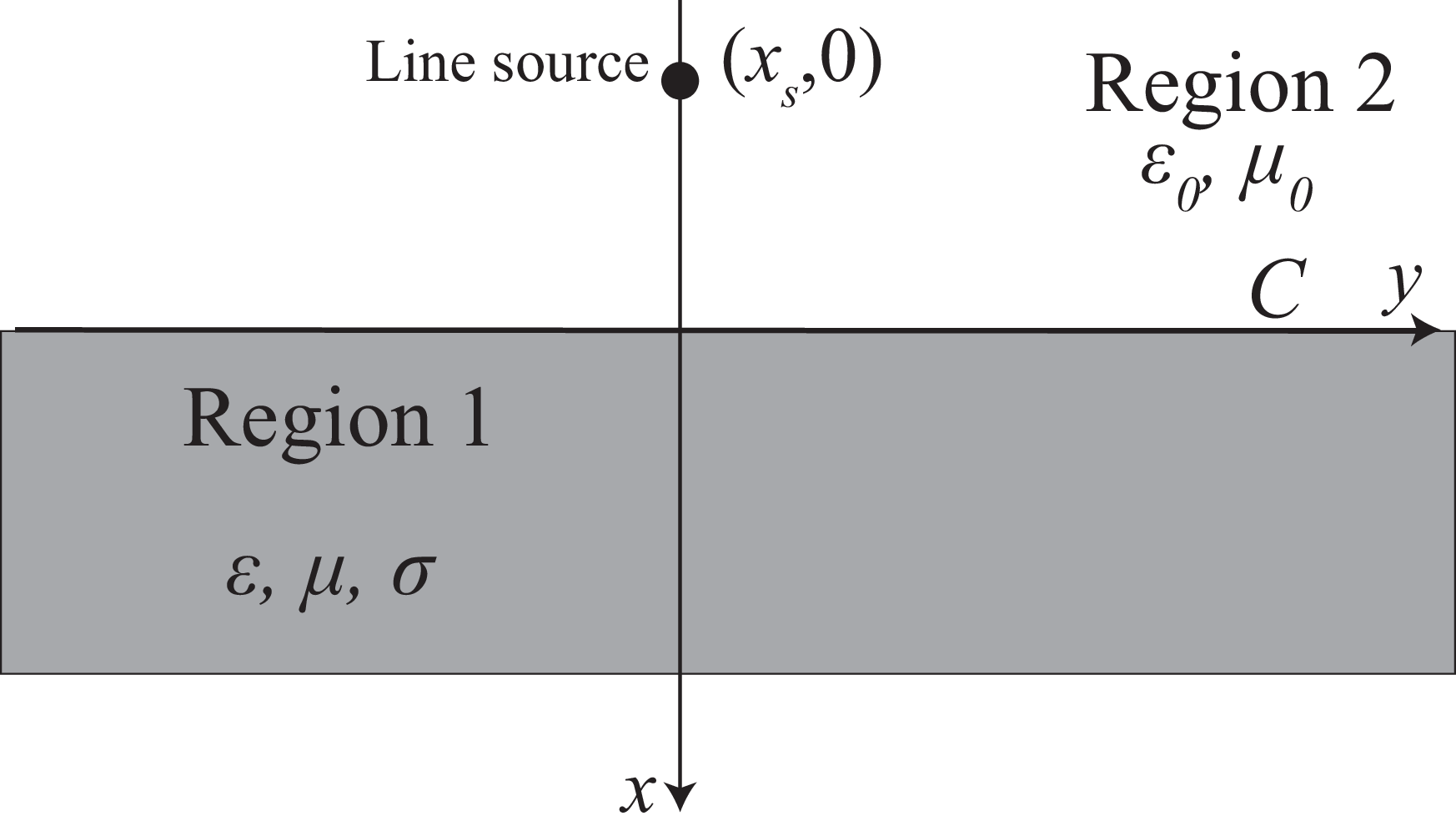}} 
    \subfigure[]{\includegraphics[width=0.48\textwidth]{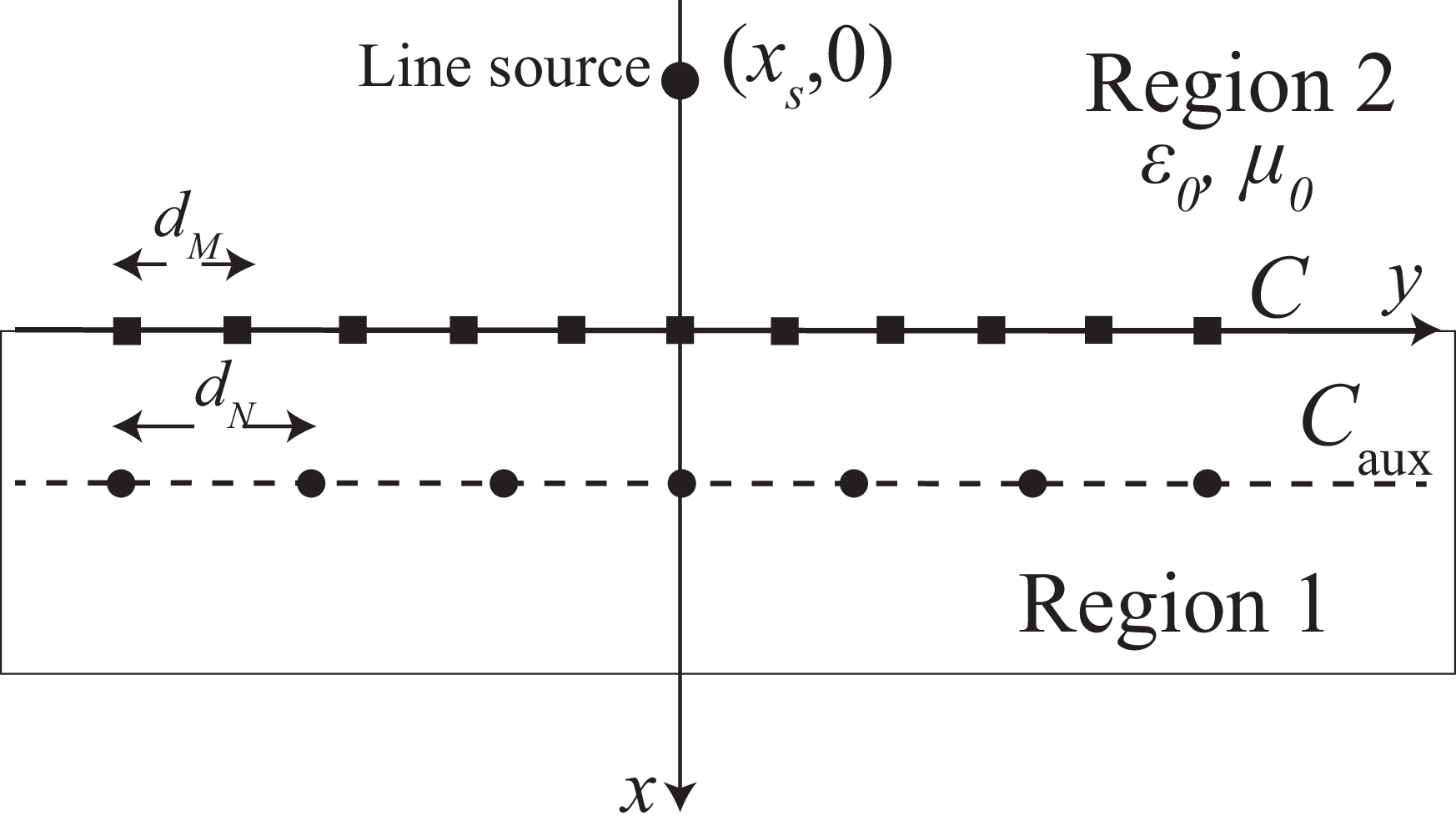}}   
    \centering
    \subfigure[]{\includegraphics[width=0.47\textwidth]{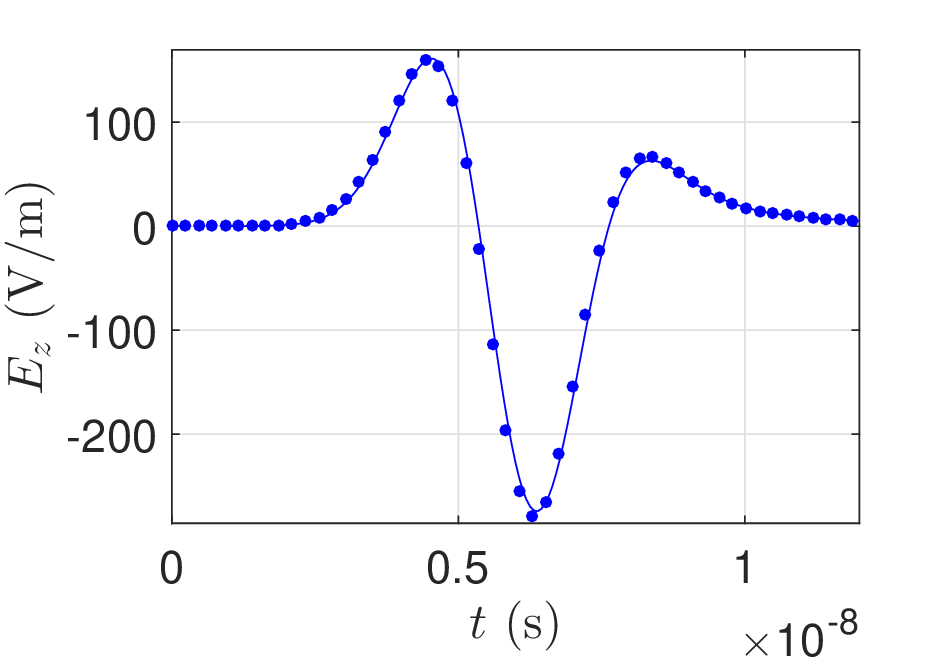}} 
    \subfigure[]{\includegraphics[width=0.47\textwidth]{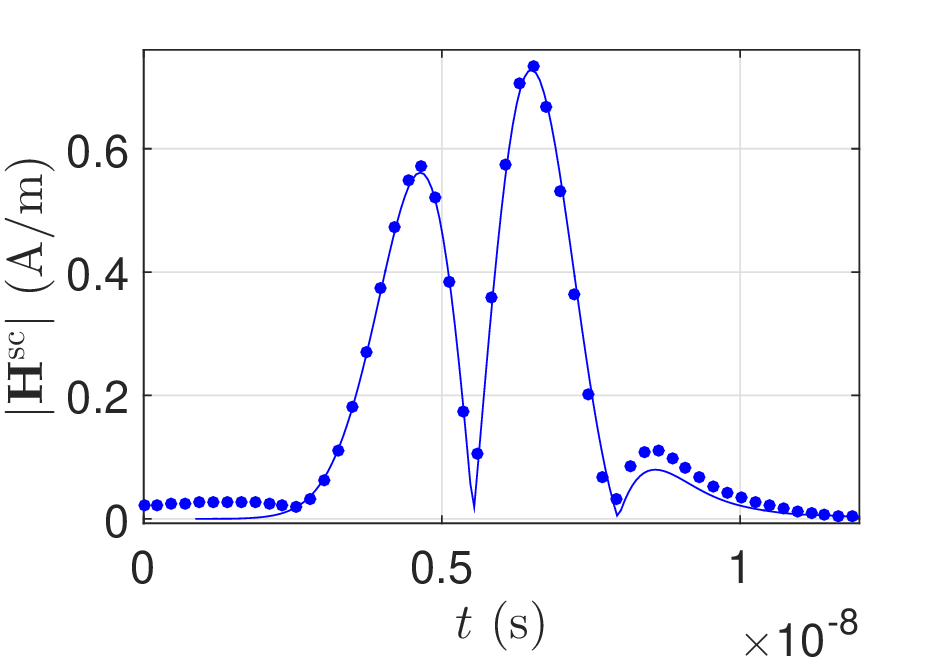}}
    \caption{(a) Schematic for the problem of line-source excitation over a lossy half plane, (b) Application of MAS, (c) Amplitude of the scattered electric field $E_z$ at the point (0, 0.3~\rm m), (d) Norm of the scattered magnetic field $|\mathbf {H}^{\rm sc}|$ at (0, 0.3~\rm m). For  (c), and (d) blue dots correspond to COMSOL results and blue solid lines to MAS results.}
    \label{fig:planar}
\end{figure}

In this last numerical experiment, we investigate the scattering of a $\rm TM_z$ cylindrical wave, originating from an electric current filament, over a semi-infinite conducting plane, as depicted in Fig.~\ref{fig:planar}(a). Since the plane is considered to be of infinite extent, periodic boundary conditions would typically be required. However, due to the nature of the excitation source---which is localized in the $xy$-plane---the radiated field decays rapidly with the distance from the source. 
This allows us to apply the MAS using a finite number of sources and matching points, as illustrated in Fig.~\ref{fig:planar}(b).

For this geometry, an odd number of matching points (\(2M+1\)) and auxiliary sources (\(2N+1\)) is used in order to achieve a symmetric distribution around the central point (the \((M+1)\)-th and \((N+1)\)-th, respectively). The optimal spacing of matching points and auxiliary sources was empirically found to be \( d_M = 1.6\frac{2\pi |x_s|}{2M+1} \) and \( d_N = 1.6\frac{2\pi |x_s|}{2N+1} \), respectively.

For the simulations, the excitation line source is placed at $(x_s, y_s)=(0, -0.3~~\rm m)$. We used $\tau=10^{-9}~\rm s$, $N = 12$, and $M=48$, $\Delta t = 0.25 \Delta t_0$. The results are presented in Figs.~\ref{fig:planar}(c) and (d), where the amplitude of the scattered electric field $E_z$ and the norm of the scattered magnetic field $|\mathbf {H}^{\rm sc}|$ are plotted at the observation point $(x_o, y_o)=(0,0.3~\rm m)$, i.e., on the planar boundary $C$. 

\section{Conclusions}

This paper presented a time-domain implementation of the Method of Auxiliary Sources (MAS) in conjunction with the Standard Impedance Boundary Condition (SIBC) for modeling electromagnetic scattering from cylindrical scatterers with moderate conductivity. The proposed method relies on a first-order SIBC formulation, which is valid under the condition that the conductivity of the scatterer significantly exceeds the product of its permittivity times the maximum spectral frequency of the incident wave. A complete theoretical framework was developed to transform the SIBC from the frequency to the time domain, and to implement it efficiently within the MAS scheme. The approach was applied to a broad range of geometries, including circular, elliptical, super-circular, rounded-triangular, and inverted-elliptical cylinders as well as planar boundaries. Simulations were performed for both TM$_z$ and TE$_z$ polarizations, under excitation by transient plane and cylindrical waves. Several values of conductivity were considered to assess the robustness of the method.

Numerical results were validated against analytical solutions and commercial frequency-domain solvers, demonstrating excellent agreement. The effect of spatio-temporal resolution was thoroughly investigated, confirming that smaller time steps and denser auxiliary source/matching point configurations lead to improved accuracy, as measured by both field comparisons and boundary error metrics. Overall, the time-domain MAS-SIBC approach proved to be an accurate alternative for modeling scattering from highly conductive objects. Future work may explore the extension of this method to more general three-dimensional configurations and the implementation of a boundary condition for thin electromagnetic shields.


\bibliographystyle{elsarticle-num}
\bibliography{TDM_MAS_SIBC}


\end{document}